%% file: delarosa.tex
\documentclass[11pt,a4paper]{emulateapj}
\usepackage{epsfig}
\usepackage{amsmath}
\usepackage{natbib}

\begin{document}
\title{Truncated Star Formation in Compact Groups of Galaxies: A Stellar Population Study}

\author{Ignacio G. de la Rosa\altaffilmark{1},\altaffilmark{2} 
Reinaldo R. de Carvalho\altaffilmark{3},
Alexandre Vazdekis\altaffilmark{1}
and
Beatriz Barbuy\altaffilmark{2}}
\date{\today}

\altaffiltext{1}{Instituto de Astrof\'\i sica de Canarias, E-38200 La Laguna, Tenerife, Spain; irosa@iac.es, vazdekis@iac.es}
\altaffiltext{2}{Instituto Astron\^omico e Geof\'\i sico - USP, 04301-904 S\~ao Paulo, Brazil; 
ignacio@astro.iag.usp.br, barbuy@astro.iag.usp.br}
\altaffiltext{3}{Instituto Nacional de Pesquisas Espaciais - INPE, DAS, S\~ao Jos\'e dos Campos - SP - Brazil; 
reinaldo@das.inpe.br}

\begin {abstract}
We present the results of a study comparing the stellar populations in the elliptical galaxies of Hickson Compact Groups (HCGs) with those in low density environments. We analyze a sample of intermediate resolution (4.25\AA~ FWHM) spectra of 22 galaxies in HCGs and 12 in the field or in loose groups. Three different population synthesis models and stellar population analyses are used to make the results more robust. Great care has been taken to correct for the emission contamination of the Balmer lines, used to determine the mean luminosity weighted ages of the galaxies. We find that, on average, galaxies in HCGs are systematically older and more metal-poor than in the field, in agreement with previous works. The most interesting finding is that the low-$\sigma$ galaxies in HCGs show an enhanced [Mg/Fe] ratio and depleted metallicity [Z/H] with respect to their counterparts in the field.  This anomalous behavior is interpreted as evidence for the action of a mechanism which truncated the star formation (SF). Hydrodynamical simulations of galaxy mergers (Di Matteo et al. 2005) support this interpretation by predicting the quenching of star formation soon after the merger event. Combining this scenario and the evidence presented here, the HCGs, generally considered to be ideal environments for galaxy-galaxy interactions, become ideal places for SF truncation.  

\end{abstract}


\section{Introduction}

In the last decade, work on Hickson Compact Groups (HCGs) has been characterized by a sense of disappointment. On one hand, HCGs of galaxies, with their high spatial densities and low velocity dispersions appear to be ideal places for interactions and mergers. On the other hand, the interaction-activity paradigm (e.g. Barnes \& Hernquist 1992) predicts the triggering of star formation and a widespread flux increase at almost every wavelength as a result of these interactions. Although observations have found abundant traces of tidal interactions in HCG galaxies (Mendes de Oliveira \& Hickson 1994), their star formation levels are surprisingly similar to those found in isolated galaxies (e.g. Zepf, Whitmore \& Levison 1991; Moles et al. 1994; Allam et al. 1996; Verdes-Montenegro et al. 1998; Iglesias-P\'aramo \& V\'\i lchez 1999). 

There is evidence that this unexpected behavior is not exclusive to the HCGs, but is shared by the broader family of galaxy groups and clusters. The exploitation of large scale redshift surveys has allowed the study of large samples of galaxies in clusters (Hashimoto et al. 1998) and groups (Wilman et al. 2005). One of the most appealing results of these works is the realization that the dense galactic environments, found in groups and clusters, are hostile to star formation. Unfortunately, the mechanism which truncates the star formation in groups remains virtually unknown. Although, there is no shortage of proposed mechanisms to halt star formation in dense environments, they are considered ineffective in group conditions. In addition, the merger of galaxies, favored by the conditions in HCGs, has also been usually excluded because it enhances the star formation instead of stopping it. 

This view has been modified based on recent hydrodynamical simulations of galaxy mergers by Di Matteo, Springel \& Hernquist (2005), which take the feedback from supermassive black holes into account. They show that the star formation enhancement is just a transient phase of a longer process which, ends up depleting the gas and quenching both star formation and nuclear activity. A short time ($\sim$ 0.5 Gyr) after the merger event, the remnant has evolved to a dead spheroid where the tracers of the transient starburst begin to fade. The short-lived SF tracers, such as the H$\alpha$ emission, disappear along with the gas, while the color, which slowly reddens after star formation has ceased, reaches u-r $\sim$ 2.2--2.3. This places the remnant in the red sequence of the bimodal color distribution in about 1-2 Gyr (Springel, Di Matteo \& Hernquist 2005). From this new perspective, the failure of HCGs to display traces of ongoing or recent star formation would not seem so disappointing. 

A more appropriate tool to determine the star formation histories (SFH) of galaxies is the study of their stellar populations. The use of this technique has produced valuable insights into the history of large samples of galaxies in high and low-density environments (Kuntschner et al. 2002; Thomas et al 2005). There also exist a few stellar population studies with smaller samples of early-type galaxies in HCGs (de la Rosa et al. 2001b; Proctor et al. 2004; Mendes de Oliveira et al. 2005). Their common result is that early-type galaxies in HCGs are older and more metal-poor than their counterparts in the field. Opposite results have also been reported (Ferreras et al. 2006). Our goal in this work is to use stellar populations to interpret the relics of the past episodes of galaxy mergers. The present study supersedes a previous work (de la Rosa et al. 2001b), using the same sample but with a less sophisticated analysis.        

The study of stellar populations in galaxies has reached maturity. It has overcome several obstacles, like the age-metallicity degeneracy, which was partially lifted with the introduction of suitable Lick spectral indices (Worthey 1994). The non-solar element abundance ratios in the galaxies was taken into account with the inclusion of the abundance ratios effects in the stellar population models (Weiss, Peletier \& Matteucci 1995; Vazdekis et al. 1997; Tantalo, Chiosi \& Bressan 1998; Trager et al. 2000a; Thomas, Maraston \& Bender 2003, hereafter TMB03; Barbuy et al. 2003). Despite these successful achievements, stellar population studies are still not problem-free. Although novel approaches and better models appear at a good pace, the convergence of their results is not accompanying these improvements. Different stellar population approaches applied to the same galaxy do not generally agree on the absolute values of the derived parameters.  Another drawback of stellar population studies arises from the relatively frequent emission contamination of the Balmer absorption indices, which are extensively used to determine galaxy ages. In the present study, the emission fill-in of the Balmer lines is a potential source of systematic errors, due to a possible different incidence of AGNs in the two environments (Coziol, Iovino \& de Carvalho, 2000). If left uncorrected, the emission fill-in weakens the Balmer absorption lines and leads us to derive systematically larger ages and lower metallicities. Some authors have abandoned emission correction of their galaxies, simply excluding the contaminated cases from their samples. 

In this work we use medium-resolution spectra of a homogeneous sample of early type galaxies belonging to both HCGs and the field or loose groups. Our goal is to study the effect of the environment on the stellar population parameters (age, metallicity and abundance ratio) of galaxies within the sample. For this reason, the absolute values of these parameters are much less relevant than their relative values. We have taken special care to double-check the influence of both the choice of a particular stellar population analysis and the emission contamination on the results. We have carried out three parallel analysis of the same data using very different approaches. We first apply the Lick procedure with the low resolution TMB03 models, including the effects of non-solar element abundance ratios. We then use the SSP models presented in Mendes de Oliveira et al. (2005, hereafter MCGB05), which also work at the Lick resolution and include non-solar abundance effects. Because the work by Mendes de Oliveira et al. (2005) contains two kinds of data relevant for this study, i.e. SSP models as well as galaxy parameters, we use the acronym MCGB05 to reference the models, and the full reference to cite the galaxy results, especially in subsection 5.2. Third is the Vazdekis (1999, hereafter Vaz99) approach, with a high resolution Stellar Population (SSP) model and a different stellar population analysis strategy, where the models are adapted to the resolution of the data, instead of the usual converse method. Our aim is not to compare the performances of the three approaches, but to gain insight into the systematic differences to assess the robustness of the results. The emission correction has also been calculated following two distinct strategies. One of them, pioneered by Gonz\'alez (1993) (hereafter G93), calculates the H$\beta$ emission contamination from the emission of the nearby [OIII]$\lambda$5007 line. The second approach, introduced by Caldwell, Rose \& Concannon (2003), estimates the H$\beta$ fill-in through several steps, involving the determination of the H$\alpha$ emission and the use of the ratios between the Balmer line strengths in nebular emission spectra (Osterbrock  1989).

This paper is organized as follows: Section 2 describes the observations and the basic reduction process. In Section 3, the emission correction is discussed. Section 4 is devoted to the extraction of the stellar population parameters with three alternative methods, whose results and correlations are shown in Section 5. In Section 6 a physical scenario is proposed to account for the results, and a summary is presented in Section 7.        

\section{Observations and Basic Reductions}
\subsection{The Data Sample}

Our sample consists of 22 bright elliptical galaxies located in the cores of HCGs (Hickson 1982) and 12 bright {\it bona fide\/} E/S0s located in the field, very loose groups or cluster outskirts, hereafter named the {\it field galaxies}. Long-slit observations were obtained during a four night run (1994 January 9--12) at the KPNO 2.1 m telescope with the GoldCam CCD spectrometer. By using a single telescope and optical configuration we guarantee the homogeneity of the sample. A 600 line/mm grating was used to yield a $\sim$ 1.25 \AA\ /pixel dispersion and a 4.25 \AA\ FWHM spectral resolution over the wavelength range 3500 to 7000 \AA. A more detailed description of the observations has been published in de la Rosa, de Carvalho \& Zepf (2001a, hereafter Paper I)).

The main details of the data have been summarized in Table 1, where column 1 shows the list of observed objects. In general, the slits were not aligned along a preferred axis of the galaxies. Repeated observations of the same galaxy were combined except for those cases where slit inclinations differed by more than 20 degrees. In those particular cases where several observations of the same object exist, each different slit inclination is marked with a plus (+) and treated as an independent object. A positive side effect of this discrimination is to increase the number of observations to populate the statistical relations. However the results of this study (Section 5) only include one observation per object, therefore, in the case of multiple inclinations only the observation with the highest S/N has been selected. 

The S/N per \AA\, shown in column 2, have been calculated for the R$_{eff}$/2 aperture and the 4795 \AA\ to 5465 \AA\ spectral interval (RED). Calculations take into account Poisson statistics, with the gain (2.8 e$^-$/ADU), readout-noise (8.5 e$^-$), number of combined spectra, and the sky value. The S/N ranges from 20 to 250, with a median value of 55. We have not excluded any galaxies in the sample based on their S/N.

\subsection{Basic Reduction and Aperture Extraction}

Data reduction was performed using the standard IRAF tasks, including bias subtraction, flat-fielding, cosmic rays and bad pixel removal, etc. The extinction correction and flux calibration of the spectra were made with the spectrophotometric standard stars Feige 34 and Hiltner 600, which were observed in each observing session.    

Aperture extraction was performed in a coherent way to allow for comparison of all galaxies in the sample and avoid the use of aperture corrections to the spectral indices. Extracted apertures are fractions of the effective radius (R$_{eff}$); therefore, we have worked out the R$_{eff}$ value along the slit during the observations. Both the data for the calculation and the results are presented in Table 1.   

Isophotal photometry data, $R_{eff}$ (column 3) and ellipticity (column 5) were taken from Zepf \& Whitmore (1993) and Trager et al. (2000b). Missing data for three of our sample galaxies (HCG 46a, HCG 51a and HCG 62a) were estimated through indirect methods, using the roughly linear relation between the de Vaucouleurs diameter ($D_{25}$) and  $R_{eff}$. Ellipticities for these three galaxies were taken at isophote 25 from the Lyon-Meudon Extragalactic Database (LEDA)

The slit inclination is presented in column 4 as the $\theta$ angle between the major axis of the ellipse and the slit. This value has been calculated from the slit orientation and the value of the major axis position angle (PA) from LEDA. Three galaxies (HCG 59b, HCG 93a and NGC 4552), with rather low ellipticity, lacked the PA information, which were directly measured from images in the NASA/Ipac Extragalactic Database (NED). The previous data are used to calculate the $R_{eff}$ value along each particular slit inclination. Column 6 shows the aperture values for $R_{eff}/8$ (in arcsecs) projected along the slit. Although only the $R_{eff}/2$ and $R_{eff}/8$ apertures are used here, smaller fractions (1/4 or 1/6) replace the $R_{eff}/8$ apertures when their sizes are of the order of 3 arcsecs, the average seeing of the observations. For instance, a $R_{eff}/6$ aperture would be preferred over a $R_{eff}/8$ one when both fall inside the same blurred bin, just because it contributes with more flux. Values for those cases are given in parentheses.

\input{tab1.tex}

\subsection{Dynamical Parameters}

The spectral lines were corrected for two main sources of broadening: the rotation and the internal velocity dispersion of the galaxy. The rotational correction is carried out along the aperture extraction, where every pixel-size slice is corrected for the rotational Doppler shift, before being stacked in the extracted spectrum. The Doppler shifts were measured with the FXCOR/IRAF task, via cross-correlation with a rest frame stellar template SAO-079251 (Paper I). The central velocity dispersions of the galaxies ($\sigma_0$) have also been measured with the procedure used in Paper I, but averaging the results of three different variants: (i) using the SAO-079251 stellar template in the full spectrum; (ii) using the best fitting SSP-Vaz99 models in the RED interval (4795 \AA\ to 5465 \AA\ ) and (iii) using the best fitting SSP-Vaz99 models in the BLUE interval (3856 \AA\ to 4476 \AA\ ). Values of $\sigma_0$ for $R_{eff}/8$ are presented in column 2 of Table 2.

\section{Emission Correction}
Residual emission in the Balmer line indices produce artificially large ages in the affected galaxies and represents a potential source of systematic errors for the present study. Extracting emission lines from an underlying absorption spectrum is especially difficult for the Balmer lines, because they coincide with the absorption lines which are our main source of information about population ages. Due to this difficulty, we have double-checked the residual emission correction by using two alternative approaches. The first strategy, proposed by G93, uses metal forbidden lines, generally independent of any underlying absorption feature, to deduce the Balmer emission components. The second strategy, proposed by Caldwell, Rose \& Concannon (2003), calculates the H$\alpha$ emission to deduce the higher order Balmer emission lines.

\begin{figure}
\epsfxsize=9cm
%\epsscale{0.6}
\epsfbox{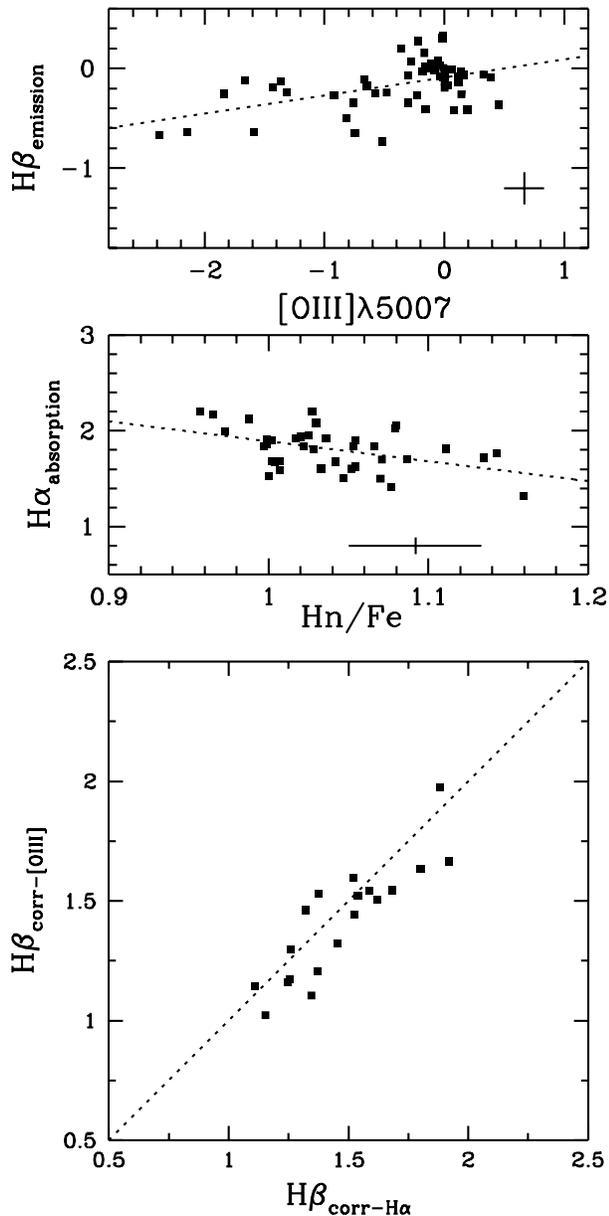}
\caption{The use of two independent methods for correction of the H$\beta$ emission. In the first method (a) the relation between the emission in [OIII]$\lambda$5007\AA\  and H$\beta$ is used to derive the contamination of the Balmer index. The second method (b) uses the relation between the {\it Rose index} Hn/Fe with H$\alpha_{absorption}$ to isolate the emission component of the H$\alpha$ line, which is used to deduce the emission contamination of the H$\beta$ line. Despite these rather scattered relations, the corrected H$\beta$ indices, with the two methods, show a good match (c). The dotted diagonal line in (c) corresponds to the equal-value line.}
\end{figure}

\subsection{First Approach}
 
The first approach seeks a relation between two spectral indices: the H$\beta_{emission}$, with the same bandpasses defined for the standard H$\beta_{absorption}$, and the nearby [OIII]$\lambda$5007\AA\  index, whose bandpasses are defined in G93. Both indices are measured in the residual spectra, which are constructed by subtracting the best fitting SSP-Vaz99 model from the original spectrum. Only data showing emission in any of the indices were used for the correlation, resulting in a linear fit of H$\beta_{emission}$ = 0.180 $\times$ [OIII]$\lambda$5007 - 0.092, with rms=0.208 (Figure 1(a)). According to this statistical correction, an [OIII]$\lambda$5007 emission of, say -0.4, implies a H$\beta$ emission filling of -0.164, i.e. a $\Delta H\beta$ = 0.164 \AA\  correction to be added to the contaminated H$\beta$ index. 

\subsection{Second Approach and Comparison}

Caldwell, Rose \& Concannon's (2003) approach is extensively explained in their paper and we will just outline the steps we have followed. In short, one measures the H$\alpha$ emission to deduce the corresponding H$\beta$ contamination. We divide our sample into two sets based on the presence or absence of [OIII]$\lambda$5007 emission. The {\it template set} is comprised of galaxies with a S/N greater than 40 and [OIII]$\lambda$5007 $>$ -- 0.3,  while the {\it contaminated set} contains galaxies with [OIII]$\lambda$5007 emission (EW $<$ -- 0.3). We have used all our R/2 and R/8 spectra as independent data, resulting in 19 {\it contaminated} and 48 {\it template} spectra. Furthermore, to improve the quality of the template set, we have excluded seven spectra with the largest departures from the mean relation. Two additional spectra were excluded due to poor spectral quality in the H$\alpha$ region. We then measure the H$\alpha$ absorption and the {\it Rose index} Hn/Fe for each template galaxy. A linear relation is found between H$\alpha_{absorption}$ and Hn/Fe (Figure 1(b)), suggesting that this last index is a reliable predictor of the H$\alpha$ absorption. Our least-squares fit gives H$\alpha_{absorption}$ = - 2.08 $\times$ (Hn/Fe) + 3.97, with a rms scatter of $\pm$ 0.19, in good agreement with the relation found by Caldwell, Rose \& Concannon (2003). We make the assumption that this same linear relation also applies to the emission-contaminated galaxies, yielding an indirect method to estimate the H$\alpha_{absorption}$ by means of the Hn/Fe index. Each contaminated galaxy is assigned a template with both a similar Hn/Fe index and a lower velocity dispersion. The previous assumption means that, after the template has been broadened to match the $\sigma$ of the contaminated galaxy, both spectra should have similar H$\alpha$ absorption. The H$\alpha$ emission can now be measured in the residual spectrum (contaminated -- $\sigma$-broadened template). Once the H$\alpha$ emission has been obtained and assuming the Balmer emission decrement ratios proposed by Osterbrock (1989), e.g. H$\alpha$/H$\beta \sim$ 3.05, the higher order Balmer emission lines can be estimated. Note that the low dust extinction in our sample galaxies alleviates the need for corrections to the theoretical decrement ratios, because they are irrelevant compared to the uncertainties of the models and the extinction curves. The intensity of emission in H$\beta$ is not yet the correction we are seeking, because Lick indices do not only measure the spectral lines after which they have been named, but the neighboring lines included in the bandpasses. An artificial emission spectrum is created with the IRAF/mk1dspec routine, including the H$\alpha$ emission line, with the flux measured in the residual spectrum, and the decremented higher order Balmer emission lines. Finally, each artificial emission spectrum is subtracted from the contaminated one and the emission-corrected Balmer indices can be measured. 

Despite the fact that both methods are completely different and based on rather scattered relations (Figures 1 (a) and(b)), there is a good match between their results. Figure 1(c) compares the emission-corrected H$\beta$ indices measured with the two alternative approaches. The comparison shows a rms scatter of 8\%, with 4\% systematically larger values of the corrected index obtained with the second approach. Translated into ages, the uncertainty and the shift correspond to roughly 2 and 1 Gyrs, using the grid of Figure 5(c). Considering the relatively good agreement between both corrections we have adopted their average as the final value. Only 19 contaminated spectra ($R_{eff}/8$ and $R_{eff}/2$) with EW[OIII]$\lambda$5007 $<$ -- 0.3, have been corrected for emission and are identified with an asterisk in Table 2.

\section{The Stellar Population Study}

There are a variety of models and strategies to calculate stellar population parameters in galaxies. We use three alternative stellar population analyses with the aim of avoiding systematic effects due to the choice of models, rather than to evaluate the performance of each analysis. We first present the standard Lick/IDS approach, with the SSP models developed by TMB03, which include abundance ratio effects. A second approach, MCGB05, uses SSP models based on synthetic stellar spectra computed in Barbuy et al. (2003). The third approach is the model/analysis proposed by Vaz99, a novel strategy which leaves the original spectral resolution of the data untouched.

\begin{figure}
\epsfxsize=9cm
\epsfbox{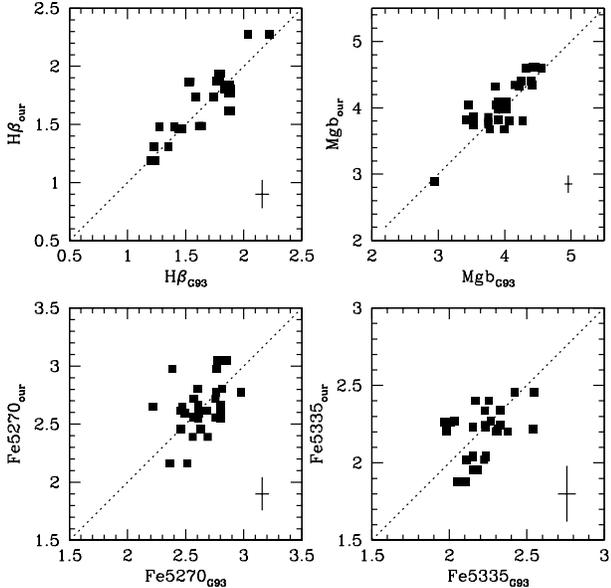}
\caption{Comparison of the raw values for the most relevant spectral indices with G93. The dotted diagonal represents the equal-value line, while the average errors are plotted in the lower right corner.}
\end{figure}
 
\begin{figure}
\epsfxsize=9cm
\epsfbox{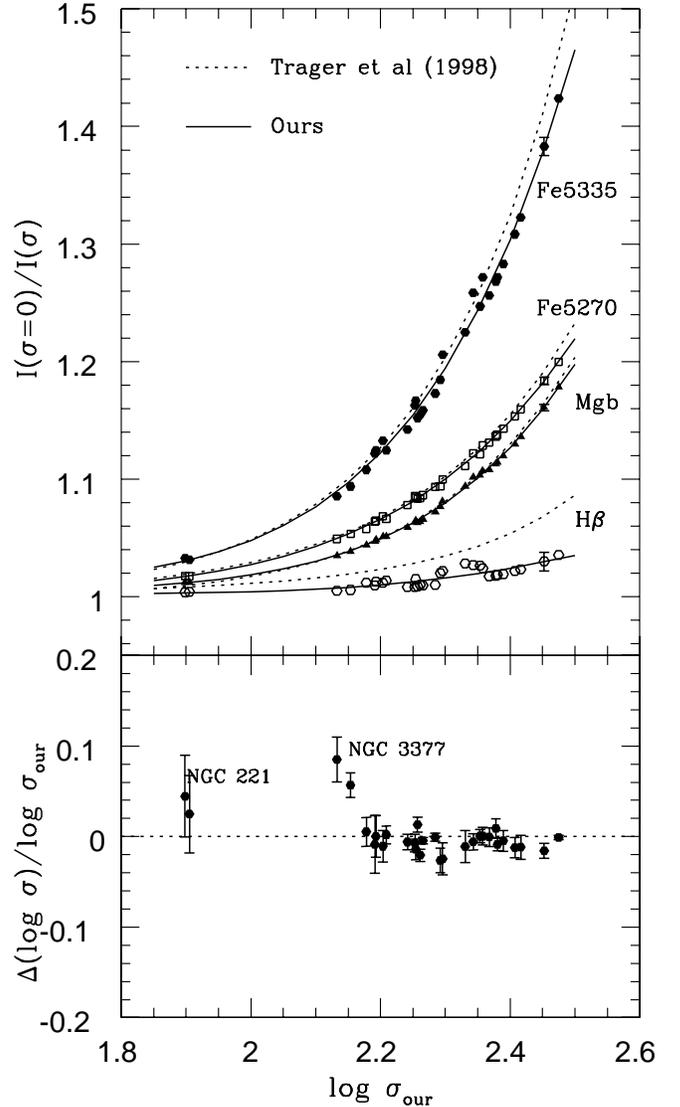}
\caption{The broadening corrections for the most relevant spectral indices  compared to those of Trager et al. (1998), showing that our values are systematically lower. Figure (b) shows the comparison, with G93, of the velocity dispersions, where $\Delta\sigma$ = $\sigma_{ours}$ -- $\sigma_{G93}$. Our values are systematically lower.}
\end{figure}
 
\begin{figure}
\epsfxsize=9cm
\epsfbox{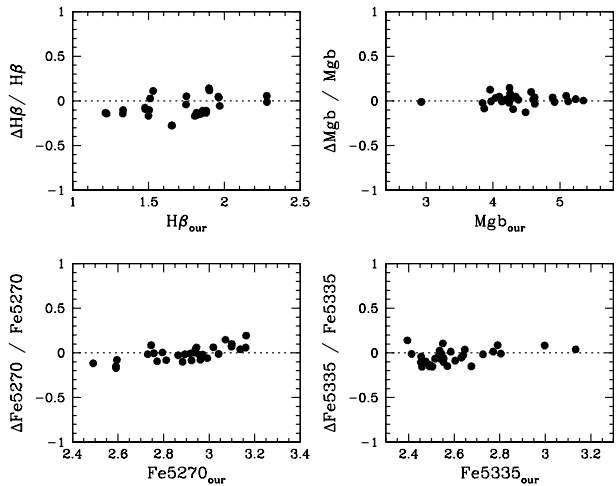}
\caption{External comparison, with G93, of the most relevant spectral indices after the broadening and emission correction have been applied. The $\Delta$Index = Index$_{ours}$ - Index$_{G93}$.}
\end{figure}
 
\subsection{The TMB03 Approach} 

\subsubsection{Transformation to Lick/IDS}
Although we lack standard stars in common with the Lick library, we have 12 galaxies in common with G93, one of the reference works in the Lick/IDS system. Comparison with those galaxies will be used to assess the quality of our transformation to the Lick/IDS system. Because G93 has data for both {\it central} $R_{eff}/8$ and  and {\it global} ($R_{eff}/2$) indices, we have a sample of 30 measurements. Our spectra were broadened to match the Lick/IDS resolution quoted by G93 (8.2 \AA\ FWHM), while his data had corrections for both the broadening and the emission removed to recover the raw index values. The index comparison shows that we have carried out a satisfactory transformation to the Lick/IDS system, with a negligible systematic discrepancy of  2\% ($<$ I$_{ours}$ - I$_{G93}>$), and a dispersion less than 8\%. Figure 2 shows this raw index comparison, with the dotted line representing equal values, and the average errors plotted.  
  
\subsubsection{Error Determination and Broadening Correction} 

The main source of error in the measurement of the line-strength indices is Poisson noise, which we calculate following the analytical approach of Cardiel et al. (1998). An almost irrelevant contribution comes from the error in the adopted zero point of the rotation curve, used to de-redshift the spectra. The redshift errors are estimated by measuring the radial velocities of each spectrum with respect to the three different rest frame templates: the stellar template and two SSP models corresponding to the RED and BLUE spectral intervals. Through this calculation, several sources of error are taken into account, including template mismatch, spectral intervals, wavelength calibration, etc. Both Poisson and redshift errors are finally quadratically added to give the total errors of the indices.

A broadening correction is applied to the raw spectral indices in order to remove the effect of the velocity dispersion of the galaxy. This correction, a feature of the Lick approach, is calculated for each galaxy and spectral index. Our templates are the best matching SSP-Vaz99 models, which can be customized for each galaxy by broadening them with both the instrumental and galaxy dispersions. For each spectral index, the broadening correction is the ratio between the value of the index in the non-broadened versus the broadened template. Figure 3(a) shows the comparison of our broadening corrections with those of Trager et al. (1998), where the largest discrepancy is found for the H$\beta$ index. A similar discrepancy was also found in the Trager et al. (1998) vs. G93 comparison.  This important multiplicative correction amplifies any error in the velocity dispersion of the galaxies, introducing a new source of uncertainty in the determination of the spectral indices. We have taken this effect into account in the final error estimate, adding it quadratically to the errors of the indices and those of the emission correction.

\subsubsection{External Comparison of the Spectral Indices}

Because this work utilizes two subsamples, a comparison to external data is unnecessary. However, a comparison with the G93 results demonstrates that despite an initial good match of the uncorrected indices, the broadening correction is a source of systematic discrepancies. This is due to its capacity for amplifying the discrepancies existing in the $\sigma_0$ determinations. As shown in Paper I, the value of $\sigma_0$ is rather sensitive to the choice of the spectral region used for its calculation. Figure 4 shows the external comparison (our results vs. G93) of the relevant spectral indices, after the broadening and emission corrections. Our corrected spectral indices are systematically lower than those of G93, except for Mgb. For instance, H$\beta$ is 10.6\% smaller, while Fe5270 is 1.4\%, with a slight systematic trend, and Fe5335 is 3.6\% lower. Considering that the broadening correction increases the index value, the observed deficit can be interpreted as an under-correction of our indices. This is a consequence of the fact that both our broadening corrections and $\sigma_0$ values are systematically lower than those of G93. The striking discrepancy between the H$\beta$ indices is due to the large disagreement between the H$\beta$ correction curves. The external comparison, ours vs. G93, of the velocity dispersion is presented in Figure 3(b). On average, the velocity dispersion of our galaxies is 2 \% smaller than those of G93, probably due to different spectral ranges used to calculate $\sigma_0$, as discussed in Paper I.  

\subsubsection{Age, Metallicity and [Mg/Fe] Determination with TMB03}    

The stellar population parameters are extracted using the models developed by TMB03, which take the abundance ratio effects into account. In order to break the age-metallicity degeneracy, we use the [MgFe]$\prime$ = (Mgb (0.72 $\times$ Fe5270 + 0.28 $\times$ Fe5335))$^{1/2}$  index, defined by TMB03, and the H$\beta$ index. The higher-order Balmer-line indices, H$\gamma$ and H$\delta$, suffer less from emission contamination, but we have avoided their use because, as reported by Korn, Maraston \& Thomas (2005), they are not only more sensitive to metallicity than the H$\beta$ index, but also significantly more sensitive to the [Mg/Fe] ratio. The stellar population parameter extraction is carried out through an iterative process. In the first step, we derive the age and total metallicity from the [MgFe]$\prime$ vs. H$\beta$ diagram with an [Mg/Fe] ratio arbitrarily set to 0.0. As shown by TMB03, both the [MgFe]$\prime$ and the H$\beta$ indices have low sensitivity to the [Mg/Fe] ratio and, consequently, these first derived ages and metallicities are generally close to their convergence values.

For the second step of the iteration, we construct a Mgb vs.$\langle Fe \rangle$ diagram, taking the previously derived age into account. From that diagram we extract the [Mg/Fe] ratio and proceed to the third step. We regenerate the [MgFe]$\prime$ vs. H$\beta$ diagram, but for the new value of the [Mg/Fe] ratio and extract new ages and metallicities. The iterative process continues until the three extracted parameters converge to a stable value. The parameter determination is performed through linear interpolation in the points of the index-index grids. A considerable fraction (18.5 \%) of the analyzed spectra show too low values of the H$\beta$ indices which place them outside the grid. For those cases, extrapolation is needed to measure the parameters, although it is worth noting that [Mg/Fe] varies slowly, with a typical rate of 0.002 dex per Gyr. An error of, say 5 Gyrs, would mean an increase/decrease of just 0.01 dex in [Mg/Fe]. Results for the three parameters and their errors, for the $R_{eff}/8$ aperture are shown in Table 2, columns 6, 7 and 8. Figures 5(a) and (b) show the galaxy positions in two representative index-index grids. Figure 5(a) corresponds to an [Mg/Fe] = 0, while Figure 5(b) corresponds to the median age of 8 Gyrs. This representation is not entirely legitimate, because the correct practice would be to plot each galaxy in the grid corresponding to its specific [Mg/Fe] and age. We note that this work  employs a 2004 version of TMB03, which includes the response functions from Korn, Maraston \& Thomas (2005) and supersedes in several aspects the original TMB03 based on Tripicco \& Bell (1995). 

\begin{figure*}
\begin{center}
\leavevmode
\epsfxsize=18cm\epsfbox{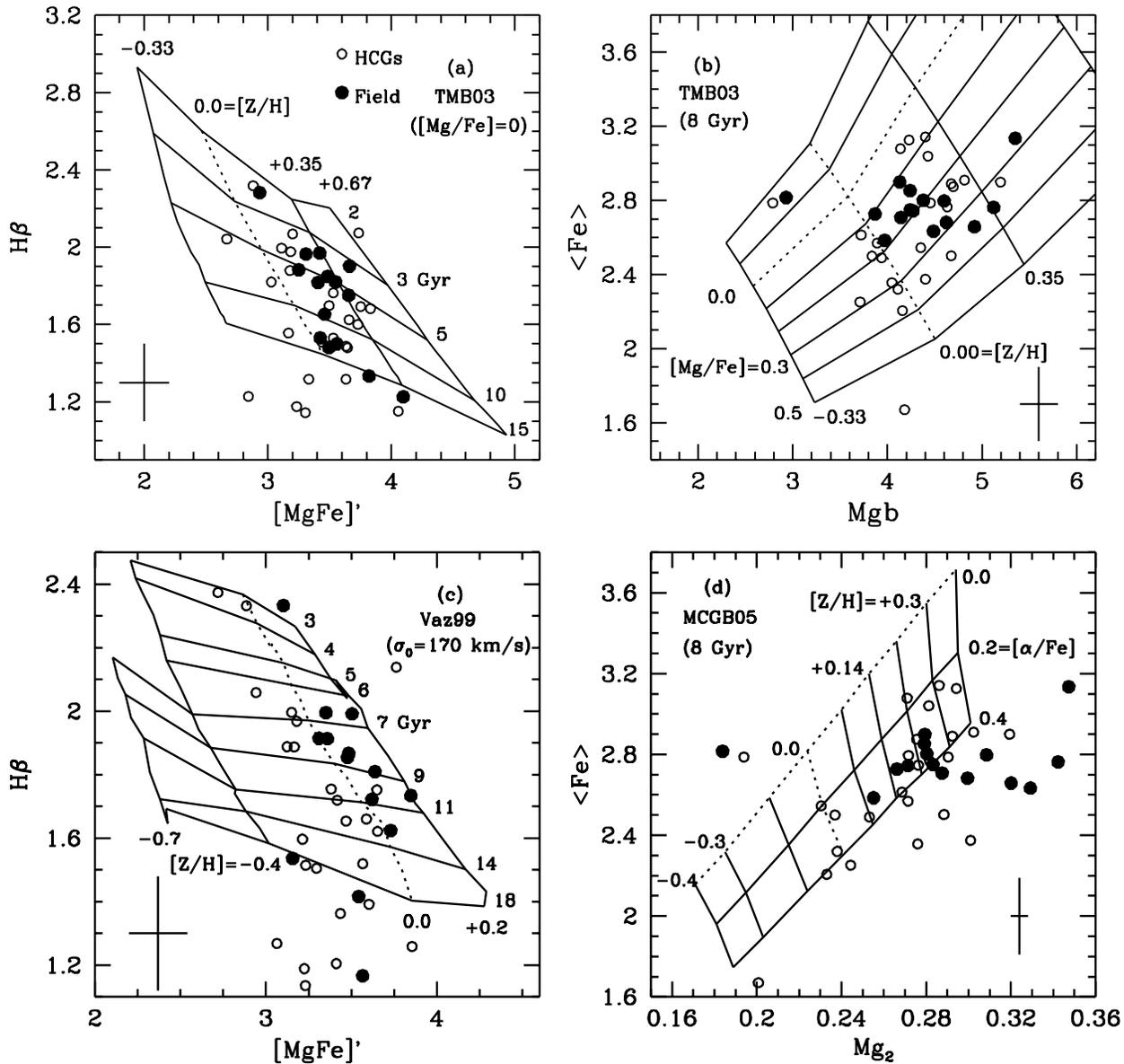}
\caption{Tools from the three approaches used to determine the stellar population parameters in the present study. Grids (a) and (b), from the TMB03 models, allow the calculation of age, [Z/H] and [Mg/Fe] through an iterative process. Grid (a) is valid for [Mg/Fe]=0.0 and  grid (b) for 8.0 Gyrs. Panel (c) shows one of the grids used in the Vaz99 analysis, to derive the age and [Z/H]. This particular grid is only valid for galaxies with $\sigma_0$ = 170 km/sec. The MCGB05 analysis uses grid (a) to calculate the ages and grid (d) to measure the [Z/H] and [Mg/Fe]. The grid represented here is valid for 8.0 Gyrs. Although all the galaxies from the field (solid) and from HCGs (open) are shown, their positions in the parameter space have no quantitative meaning, because these grids are only valid for particular values of some parameters. The proper procedure, used in this study, requires the construction of customized grids for each galaxy.}
\end{center}
\end{figure*}

\subsubsection{External Comparison of the Parameters}   

We have ten galaxies in common with Thomas et al. (2005), whose spectral indices were extracted from G93 by these authors. The differences with their population parameters are a natural consequence of the above discussed discrepancies between the spectral indices (see subsection 4.1.3). While their ages are, on average, 1.7 Gyrs lower than ours, their metallicities are larger by 0.14 dex. For the [Mg/Fe] ratios, their larger $\langle Fe \rangle$ values and smaller age conspire to produce a negligible discrepancy with our values, well within the errors of the parameter. We have just three galaxies in common with Mendes de Oliveira et al. (2005), whose Fe5270, Fe5335 and Mgb indices are up to 4\% larger than ours. Due to the use of the same stellar population analysis and TMB03 models, we expect the same pattern of parameter discrepancies found with Thomas et al. (2005).

\subsection{Age, Metallicity and [Mg/Fe] Determination: The MCGB05 Approach} 

The MCGB05 approach is based on the grid of synthetic stellar spectra computed by Barbuy et al. (2003), which spans the range 4600-5600 \AA\ for stellar parameters 4000 $\leq$ T$_{eff}$ $\leq$ 7000 K, 5.0 $\leq$ log g $\leq$ 0.0, --3.0 $\leq$ 0.3, and two values of chemical abundance [$\alpha$/Fe], 0.0 and +0.4 dex (the $\alpha$-elements considered are O, Mg, Si, S, Ar, Ca, Ne, Ti). MCGB05 present SSP models for the indices Mg2, Fe5270 and Fe5335, computed with improved coefficients for the fitting functions given by Barbuy et al. (2003), and add a new set of spectra with [$\alpha$/Fe] = +0.2. In the MCGB05 approach, the [Z/H] and [Mg/Fe] parameters are calculated by interpolation (or extrapolation) of the Mg2 vs.$\langle Fe \rangle$ diagram corresponding to the age of the galaxy. Age values are taken from the prior TMB03 approach, because no age-sensitive lines were computed in Barbuy et al. (2003).

Although the TMB03 and MCGB05 approaches are apparently equivalent in some aspects, they show significant differences in their handling of the [$\alpha$/Fe] abundance ratio. TMB03 corrects the empirical fitting functions of Worthey et al. (1994) with the response functions of Korn, Maraston \& Thomas (2005) to account for the [$\alpha$/Fe] enhancements, while the SSP models of MCGB05 are based on synthetic stellar spectra whose [$\alpha$/Fe] dependence is explicitly established along with the model computation. The results from this approach are presented in Table 2 (columns 9 and 10). Figure 5(d) shows a typical metal-metal grid used for the [Z/H] and [Mg/Fe] calculations. It is worth mentioning that the superposition of the whole sample serves only for qualitative purposes, as the grid is strictly valid only for those objects with ages of 8 Gyr.

\subsection{The Vaz99 Approach}

The main advantage of the Vaz99 approach is that the high resolution SSP models are adapted to the data, instead of the other way around. Therefore, there is no need for matching or broadening corrections. Both the $\sigma_{instrumental}$ and the $\sigma_{galaxy}$ are taken into account by comparing the observed spectra with the $\sigma$-broadened SSP models. As a drawback of this approach, non-solar [Mg/Fe] abundance ratios are not incorporated in the models. As with the previous approaches, it is equally important to carry out an accurate emission correction and, due to the different spectral resolution of the data, we have repeated the double calculation of the correction for the H$\beta$ index, following the procedures of Section 3.

\subsubsection{Age and Metallicity}      

The Vaz99 SSP models have been constructed from the Jones' (1999) empirical stellar library, with high spectral resolution (1.8 \AA\ FWHM). They predict the Spectral Energy Distributions (SED) of a range of age and metallicity combinations, covering the intervals --0.7  $ \leq log(Z/Z_{\odot}) \leq $ +0.2 and 1 to 17 Gyrs. These SEDs span just a BLUE (3856 -- 4476 \AA\  ) and a RED ( 4795 -- 5465 \AA\ ) spectral interval. 

The Vaz99 stellar population analysis, which is markedly different from TMB03, proceeds as follows: first, the SSP model spectra are broadened from their original resolution (1.8 \AA\ FWHM) to the total resolution of the spectrum of each particular galaxy. This $\sigma_{tot}$ resolution is the quadratic sum of our instrumental broadening $\sigma_{ins}$ $\approx$ 133 km s$^{-1}$ and the galactic velocity dispersion $\sigma_{gal}$ taken from Table 2 (column 2). Once the broadened SSP models match the resolution of each particular galaxy, the comparison proceeds by measuring the spectral indices of the whole SSP model set. Our observed indices are then superimposed on the H$\beta$ vs. [MgFe]' grid, constructed with the SSP values. The age and metallicity parameters are estimated by interpolating among the SSP grid points. The SSP model grid is only valid for each $\sigma_{gal}$, given that $\sigma_{ins}$ is a constant for the whole sample. Although it is not the case for the present study, it would be also possible to smooth a sample of galaxies to the same $\sigma_{tot}$, in order to allow the use of a common grid. An extrapolation of the SSP model grid is sometimes needed to account for metallicities larger than the model upper limit [Z/H] = +0.2. Results for the central aperture $R_{eff}/8$ are presented in Table 2, columns 3, 4 and 5.  

Errors for ages and metallicities are also estimated with the grid. The error bars run from the locus of maximum to minimum index values. Although the error bars of the spectral indices are symmetric around their central values, this is not the case for the age-metallicity parameters, because the SSP model grid is neither orthogonal nor equally spaced. 

\subsubsection{Relative Abundances}

In the approach followed by Vaz99, one measures the abundance ratio of the galaxy relative to the scale-solar one, calculating the Mg and Fe abundances separately (see e.g. Vazdekis, Trujillo \& Yamada 2004). According to Tripicco \& Bell (1995), the Mgb spectral index is predominantly sensitive to the $\alpha$ element Mg, while the combined index Fe3 is sensitive to Fe. Both the [Z$_{Mg}$/H] and [Z$_{Fe}$/H] abundances are extracted from the H$\beta$ vs. Mgb and H$\beta$ vs. Fe3 diagrams and the [Mg/Fe] ratio is ([Z$_{Mg}$/H] -- [Z$_{Fe}$/H]). The Vaz99 simple approach is hampered by inaccuracies in the [Z$_{Mg}$/H] determination, which generally requires extrapolation of the model grid for values larger than the upper limit [Z/H] = + 0.2, especially for the most massive galaxies. Furthermore, the small departures from orthogonality in the H$\beta$ vs. metal-index grids transform errors in age into errors in metallicity. They have been incorporated into the errors obtained  for the [Mg/Fe] ratios.

\subsection{Comparison of the Three Approaches}

It is worth making a comparison of the extracted parameters in order to evaluate the systematic effects introduced by the choice of the model and stellar population analysis. This is also interesting, outside the context of the present study, in order to better interpret other stellar population results obtained using any of these different approaches. In the present study, the TMB03 approach has been chosen as the common reference for comparison of all our results, for the sole reason that it makes comparison with the literature easier, as explained in section 5.2. Figure 6 shows the comparison with Vaz99, while Figure 7 shows the comparison with MCGB05.

\begin{figure*}
  \begin{center}
    \leavevmode
      \epsfxsize=18cm\epsfbox{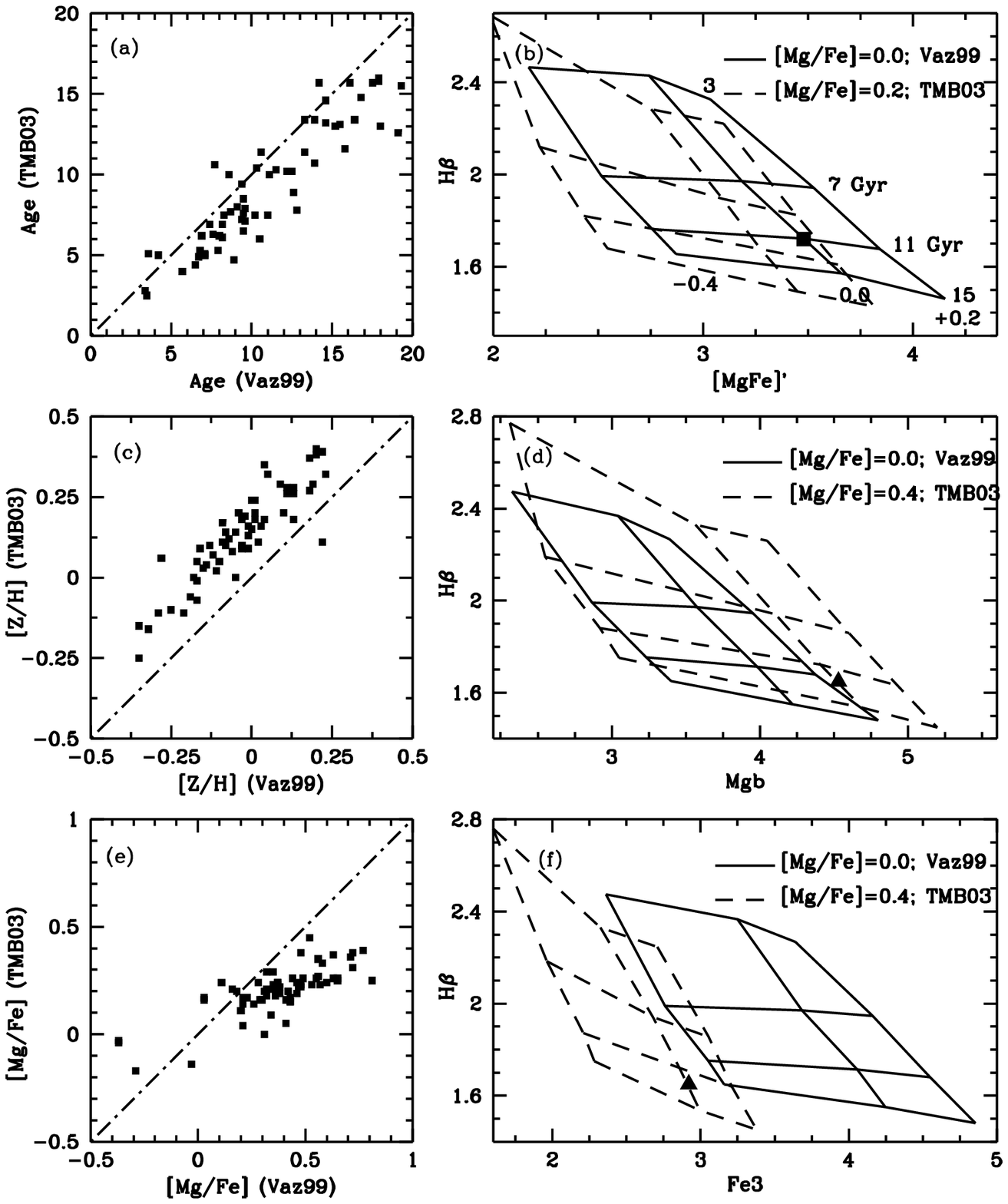}
\caption{Comparison of the stellar population parameters obtained with the TMB03 and Vaz99 models. Panels (a), (c) and (e) show the comparison of ages, [Z/H] and [Mg/Fe] for the whole sample, where systematic discrepancies are evident. Panels on the right provide qualitative explanations for the discrepancies.  In figure (b), equivalent grids from both models have been superimposed. For the same pair of spectral indices, an arbitrary galaxy (the solid square) shows systematically larger age and lower [Z/H] with the Vaz99-grid (continuous line) than with the TMB03-grid (dashed line), in agreement with panels (a) and (c). The systematic discrepancy in [Mg/Fe] is explained in figures (d) and (f) for an arbitrary galaxy (solid triangle) with [Mg/Fe](TMB03) = 0.4, i.e. a galaxy with the same [Mg/H]=[Fe/H]=0.0 in the [Mg/Fe]=0.4 grids of the TMB03 models (dashed lines). The Vaz99 procedure to measure [Mg/Fe], i.e. [Mg/Fe]=[Mg/H]-[Fe/H], gives a much larger value ([Mg/Fe] = 0.26 -- (--0.50) = 0.76), as observed in panel (e).}
\end{center}
\end{figure*}

\begin{figure*}
\begin{center}
    \leavevmode
      \epsfxsize=18cm\epsfbox{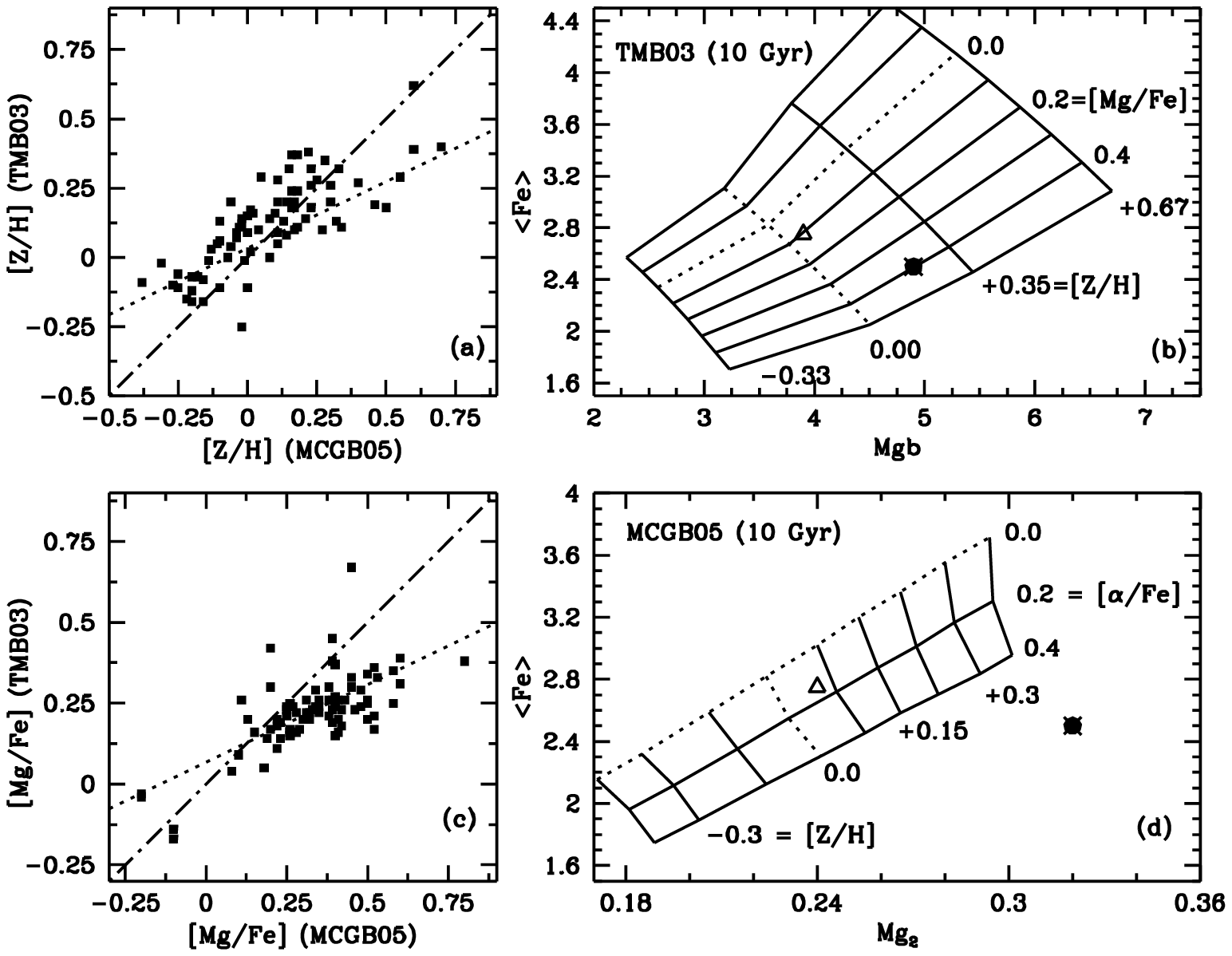}
\caption{Comparison of stellar population parameters obtained with TMB03 and MCGB05 models. Left panels show the comparison between [Z/H] and [Mg/Fe] with the equal-value line (dot-dashed) and the least-squares fit (dotted) to show the systematic discrepancies. Panels (b) and (d) explain qualitatively the discrepancies, by showing two arbitrary galaxies with [Mg/Fe](TMB03) = 0.1 (open triangle) and with [Mg/Fe](TMB03) = 0.4 (solid circle). These same galaxies, when studied with the MCGB05 models, have [Mg/Fe](MCGB05) = 0.1 and 0.66, respectively, as observed in (c). The [Z/H] shows a smaller discrepancy, as seen in (a).}
\end{center}
\end{figure*}

\subsubsection{The Vaz99 vs. TMB03 results}

As shown in Figures 6, ages obtained with the TMB03 analysis are, on average, 16\% lower than those obtained with the Vaz99 approach, while TMB03 metallicities are 0.16 dex larger, on average, than those of Vaz99. These discrepancies can be qualitatively explained by differences between the TMB03 and Vaz99 model grids, as shown in Figure 6(b). In order to make the comparison easier, all the grids of Figure 6 have been adapted to show the same age and metallicity points. Note that it is inaccurate to represent both TMB03 and Vaz99 grids in a common graph, except for the special case where the observed $\sigma_{tot}$ = ($\sigma_{gal}^2$ + $\sigma_{inst}^2$)$^{1/2}$ equals the instrumental broadening of $\sigma_{Lick}$ = 200 km s$^{-1}$. For our particular $\sigma_{inst}$, this means that both grids are roughly equivalent when $\sigma_{gal}$ = 172 km s$^{-1}$. In Figure 6(b), we have used the Vaz99 grid for a $\sigma_{gal}$ = 172 km s$^{-1}$ galaxy and the TMB03 grid with [Mg/Fe] = 0.2, the average value of the sample.  From Figure 6(b) we can understand qualitatively the observed age and metallicity discrepancies. For instance, a galaxy (solid square) with $\sigma_{gal}$ = 172 km s$^{-1}$ , H$\beta$ = 1.72 and [MgFe]$\prime$= 3.48 would have an Age$_{Vaz99}$ = 11.0 Gyrs and an Age$_{TMB03}$ = 9.0 Gyrs, as well as [Z/H]$_{Vaz99}$ = 0.0 and [Z/H]$_{TMB03}$ = 0.17. The relatively lower dispersion of the points in Figure 6(c) with respect to (a) is probably due to the fact that lines with equal-metallicity are more parallel than lines with equal-age.

Concerning the abundance ratios, the [Mg/Fe] obtained with the Vaz99 approach are larger than the [Mg/Fe] of TMB03. Furthermore, the discrepancy increases with the overabundance. These differences might be attributed to the fact that, according to Korn, Maraston \& Thomas (2005), Mg and Fe indices show inverse sensitivities to [Mg/Fe], i.e. the Mgb index increases with [Mg/Fe], while the Fe3 index decreases. As these responses are used in TMB03 models, the Vaz99 approach, which is scaled-solar, consequently overestimates [Mg/H] and underestimates [Fe/H] metallicities with respect to TMB03, leading to systematically larger [Mg/Fe] values. A similar graphic explanation of the discrepancy is presented in Figures 6(d) and (f), which show the H$\beta$ vs. Mgb and H$\beta$ vs. Fe3 grids for [Mg/Fe] = 0.0 (Vaz99) and [Mg/Fe] =  0.4 (TMB03). In the Vaz99 approach, both the [Mg/H] and the [Fe/H] metallicities are measured with respect to the [Mg/Fe] = 0.0 grid (solid-line). For any [Mg/Fe] $\geq$ 0, the Vaz99 approach overestimates [Mg/H] (Figure 6(d)) and underestimates [Fe/H] (Figure 6(f)), compared to TMB03, and leads to the observed larger [Mg/Fe](Vaz99) values. In order to quantify the discrepancy, we have represented an arbitrary galaxy (solid triangle) with $\sigma_{gal}$ = 172 km s$^{-1}$ in Figures 6(d) and (f). According to TMB03, this galaxy would have [Mg/Fe] = 0.4, because both [Fe/H] and [Mg/H] metallicities converge to the same value (0.0) only for this particular [Mg/Fe] ratio. In the Vaz99 approach, the same galaxy would have [Mg/Fe] = 0.76, because [Mg/H]= 0.26 (Figure 6(d)) and [Fe/H]= - 0.50 (Figure 6(f)). This is the typical disagreement observed in Figure 6(e) for the objects with the largest [Mg/Fe] ratios.

\subsubsection{The MCGB05 vs. TMB03 results}

Comparison of MCGB05 with TMB03 results, presented in Figures 7 (a) and (c) shows that the MCGB05 metallicities are, on average, 0.07 dex lower than TMB03. The [Mg/Fe] ratio is systematically larger in MCGB05, with the discrepancy increasing with the abundance ratio. The comparison is now easier to interpret because both methods share the same age (TMB03) values. The Mg vs. Fe diagrams for each approach are presented in Figures 7(b) and 7(d), where a common age (10 Gyr) has been selected. Two arbitrary galaxies have been represented and their [Z/H] and [Mg/Fe] parameters have been extracted through inter- or extrapolation of the grid values. In order to guarantee the equivalence of the galaxies, we have matched the two Mg indices with the relation  Mg$_b$ = 12.4 $\times$ Mg$_2$ + 0.91. This is a low-dispersion relation obtained empirically from our data, including 80 different measurements. The first example, the empty triangle, with [Z/H]$_{TMB03}$ = +0.05 and [Mg/Fe]$_{TMB03}$ = 0.10 shows comparable parameter values in MCGB05, with a slightly lower [Z/H] (Figure 7(d)), while the second one, the solid circle, with [Z/H]$_{TMB03}$ = +0.23 and [Mg/Fe]$_{TMB03}$ = 0.40 shows an inflated abundance ratio (0.66), as expected. 

In order to provide a common result for the three approaches, the Vaz99 and the MCGB05 partial results have been transformed into the TMB03 common frame of reference, by correcting their mutual systematic discrepancies. From the linear fits obtained from Figures 6 and 7 we get:

\begin{eqnarray*}
Age_{TMB03} & = & 0.87 \times Age_{Vaz99} - 0.27 \\
\left[ Z/H \right] _{TMB03} & = & 0.75 \times \left[ Z/H \right]_{Vaz99} + 0.17 \\
\left[ Mg/Fe \right]_{TMB03} & = & 0.33 \times \left[ Mg/Fe \right]_{Vaz99} + 0.10   
\end{eqnarray*}

\begin{eqnarray*}
\left[ Z/H \right] _{TMB03} & = & 0.56 \times \left[ Z/H \right]_{MCGB05} + 0.07 \\
\left[ Mg/Fe \right]_{TMB03} & = & 0.48 \times \left[ Mg/Fe \right]_{MCGB05} + 0.07   
\end{eqnarray*} 
\\
   
The final results of the present study, included in Table 2 (columns 11, 12 and 13), are an average of the individual results from the three approaches, after they have been converted to the common TMB03 frame. The final errors are computed from the errors of the individual approaches added in quadrature.

From Figures 6(c)(e) and 7(a)(c) it is easy to conclude that the metallicity parameters from the Vaz99 and MCGB05 models are in better mutual agreement than those of TMB03. This provides a sound consistency test for these two widely different approaches (Vaz99 and MCGB05).  
\section{RESULTS}

We expect the ages, metallicities and [Mg/Fe] abundance ratios of the galaxies to vary with both the internal and environmental characteristics. We have considered the galaxy mass, or its related observable the central velocity dispersion $\sigma_0$, as the most appropriate internal parameter. Concerning the environment, although our study is based in the simplistic division into HCG and the field, we have also explored the diversity of the heterogeneous HCG environment by means of the {\it crossing time} which characterize their dynamical evolution. Observations of the same galaxy with different slit inclinations are not treated as different objects; instead only the observation with the highest S/N is considered.

\clearpage

\input{tab2.tex}

\clearpage

\subsection{Stellar Population Parameters vs. Velocity Dispersion}

The observable parameter $\sigma_0$ can be indirectly linked to the stellar mass through the mass-to-light ratios and the Faber-Jackson relation (e.g. equation (2) in Thomas et al, 2005). In Figure 8 we show the correlation of the stellar population parameters with $\sigma_0$ for the two different environments. Data from the literature have also been included in the graphs and used to calculate the least-squares fits, as described in subsection 5.2. The dwarf galaxy HCG 37e has been excluded from the fits of Figures 8(c) and 8(f) to avoid its excessive statistical weight. Although only our {\it average} results are presented in Figure 8, the same qualitative behavior has been consistently found using the results from each of our three alternative approaches (see Table 2).

\begin{figure*}
\begin{center}
\leavevmode
\epsfxsize=18cm\epsfbox{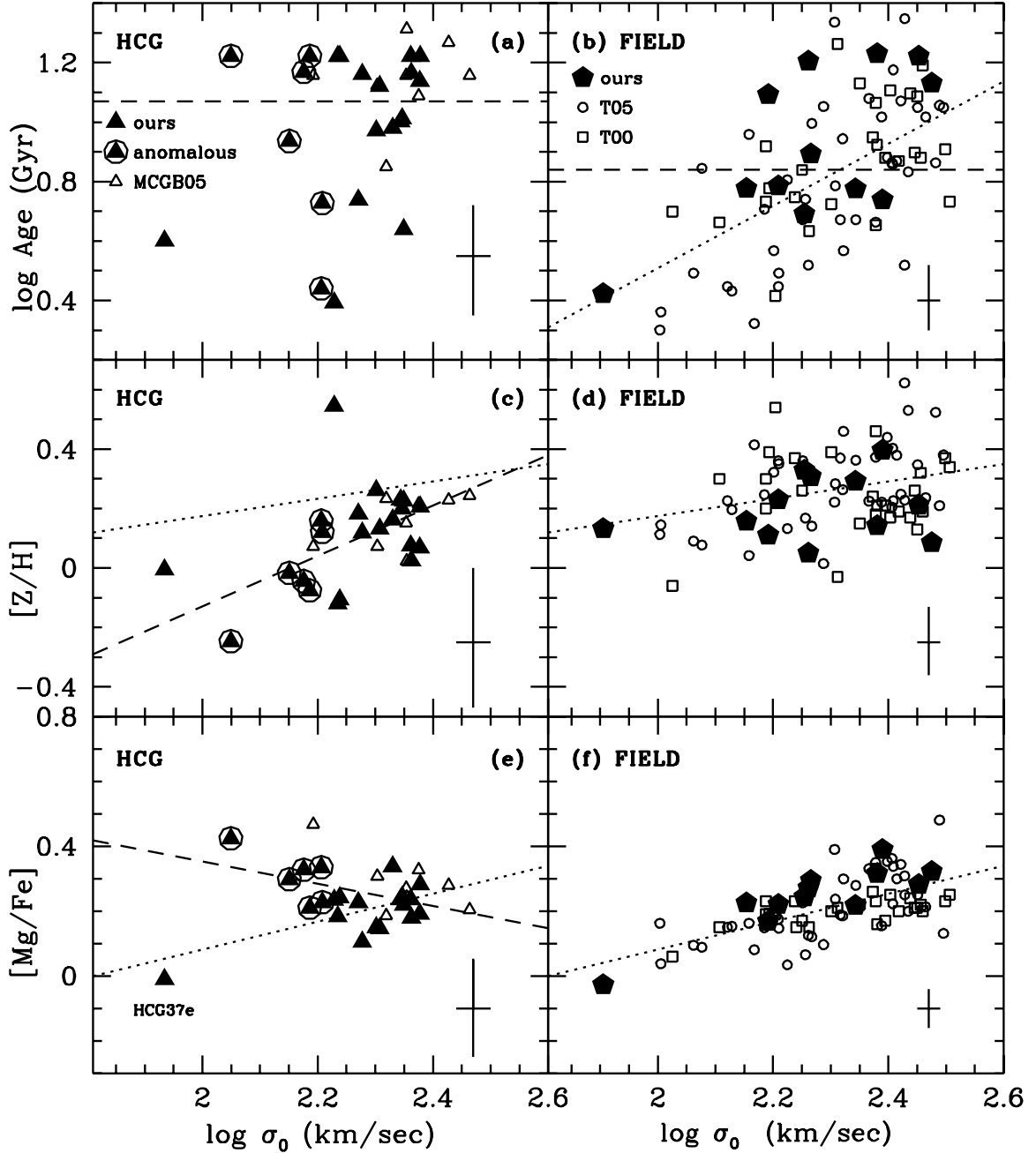}
\caption{Correlations of our results with log$\sigma_0$, comparing the behavior of elliptical galaxies in the FIELD (right-hand panels) with their counterparts in Hickson Compact Groups (HCG) (left-hand panels). Data from the literature (open symbols) have been superimposed to assess our results. They were taken from Thomas et al. (2005, T05), Trager et al. (2000a, T00) and Mendes de Oliveira et al. (2005, MCGB05) and previously converted to our reference frame. Least-squares fits (dotted lines) for all the field galaxies are displayed in figures (b), (d) and (f) and repeated in the group galaxy panels (c) and (e). Dashed lines in figures (a) and (b) show the median value of our data, while panels (c) and (e) show the least-squares fit (dashed) to all the objects, excluding the galaxy HCG37e, with log$\sigma_0$=1.93. The {\it anomalous} galaxies (see text) are tagged with a circle and the average error bars of our data are plotted in the low-right corner of each panel.} 
\end{center}
\end{figure*}

Several of the trends found in Figure 8 deserve comment. Although our data do not show significant correlation between age and velocity dispersion (the solid symbols in Figure 8(a) and (b)), they follow the large-dispersion trend for {\it field} data noted in the literature (Trager et al. 2000a; Thomas et al. 2005), and in agreement with the extensive work of Gallazzi et al. (2006) using SDSS early-type galaxies. Both the fit to all the data (dotted line) and the median value of only our results (dashed line) are shown in Figure 8(b). For galaxies in HCGs (Figure 8(a)), ages do not show any trend with log $\sigma_0$ and only the median value of our results is plotted.  There is a positive correlation between the [Z/H] metallicity and log $\sigma_0$ for the galaxies in both environments (Figure 8(c) and (d)), which is remarkably steeper in the groups. A positive correlation is generally interpreted as the action of deeper potential wells hindering the loss of metal-enriched gas through galactic winds (Greggio 1997). We interpret the steeper trend of the [Z/H] vs. log $\sigma_0$ in the group environment as the action of a SF truncation mechanism which, nevertheless, does not preclude the aforementioned processes.

The field galaxies show a clear correlation between [Mg/Fe] and log $\sigma_0$ (Figure 8(f)), in good agreement with the data from the literature. This trend has been interpreted as a sign of {\it downsizing}, meaning that the more massive galaxies complete their star formation faster (e.g. Thomas et al 2005). A novel result of the present study is the {\it negative correlation found between [Mg/Fe] and log $\sigma_0$ of the low-mass galaxies of HCGs} in contrast with the behavior of the corresponding field galaxies. Figure 8(e) shows the two contrasting linear fits, with a dashed line for the HCGs and a dotted one for the field. Despite the striking visual contrast between both trends, the size of the [Mg/Fe] errors make the difference between slopes barely significant, just slightly above the one-$\sigma$ level. Nevertheless, other peculiar behaviors reinforce the reality of the discrepancy, like the deficient metallicity of low-mass galaxies in Figure 8(c) and other photometric and environmental parameters discussed below. We have arbitrarily selected a subsample, including the low-mass HCG galaxies with 2.0 $\leq$ log$\sigma_0$ $\leq$ 2.2, because they are largely responsible for the discrepancy between the slopes of the fits in Figure 8(e). At the same time, members of this low-mass subsample, marked with circles in Figure 8, also populate a region of the parameter space where [Mg/Fe] $\geq$ 0.21 and [Z/H] $\leq$ 0.16. Hereafter, we call them {\it anomalous} galaxies, noting that this anomaly refers only to their discrepancy with the field galaxies. Dwarf galaxy HCG37e is excluded from the subsample, as its behavior coincides with that of the field galaxies. In agreement with the {\it downsizing} concept, this galaxy presents parameter values comparable to the Sun, suggesting an extended SF history, similar to our galaxy disk. 

There are some remaining sources of concerns regarding the reality of the anomalous enhancement of [Mg/Fe] in the low-mass HCG galaxies. Among them, first is the possible bias introduced by the emission correction. As explained in subsection 5.4.1, the {\it anomalous} galaxies are preferentially AGNs, i.e. they are contaminated by emission. Errors in the emission correction are propagated unevenly to the stellar population parameters. Their sensitivity to changes in H$\beta$ is largest for the age and lowest for the [Mg/Fe]. For instance, an unrealistically large error in the correction of $\Delta$H$\beta\sim$ 0.3 is converted into $\Delta$age $\sim$ -- 7 Gyrs, $\Delta$[Z/H] $\sim$ 0.2 dex and $\Delta$[Mg/Fe] $\sim$ 0.02. Figures 5(a) or 5(c) are useful to understand the qualitative effect of a change in H$\beta$. Only a scenario in which the emission-contaminated spectra of the low-mass galaxies are systematically under-corrected by an exaggerated amount would reproduce the behavior observed in the [Z/H] vs. log $\sigma_0$ trend of the  HCG galaxies, although it would leave the {\it anomalous} trend in [Mg/Fe] vs. log $\sigma_0$ almost unaffected. Considering the double emission correction carried out in section 3, we have no reason to suspect such a tweaked under-correction.  

A further source of concern is that this anomalous effect has passed unnoticed in previous studies, i.e. Proctor et al. (2004) and Mendes de Oliveira et al. (2005). As we show in the next subsection, their results generally agree with ours, but the shortage of data for low-mass galaxies, in the interval log$\sigma_0$ = 2.0 - 2.2, prevented these authors from observing the effect.  

\subsection{Data from the Literature}

Several other studies on the stellar populations of early-type galaxies can be used to assess the correlations found here and to expand our samples. For field galaxies, we have used the works of Thomas et al. (2005) and Trager et al. (2000a). Least-squares fits to the correlations of the stellar population parameters, i.e. Age, [Z/H] and [Mg/Fe], with log$\sigma_0$ have been calculated for our data and the external samples. Despite slight systematic discrepancies, explained below, the average agreement among the three samples is always within the rms deviations around the fits, demonstrating good agreement of our field galaxy results with the literature. 

In addition to providing indpendent comparisons, the works of Thomas et al. (2005) and Trager et al. (2000a) can be used to expand our sample, because they share several galaxies in common with our field subsample, which allow for conversion to the TMB03 reference frame of our results. Galaxies from the "low-density" subsample of Thomas et al. (2005) have been corrected to our reference frame via 12 common galaxies. The conversions for the stellar-population parameters, a measure of the systematic discrepancies, are Age$_{corr}$ = 1.07 $\times$ Age$_{Thomas}$ + 1.24; [Z/H]$_{corr}$ = 0.69 $\times$ [Z/H]$_{Thomas}$ - 0.02 and [Mg/Fe]$_{corr}$ = 1.1 $\times$ [Mg/Fe]$_{Thomas}$. After excluding the galaxies which are redundant with the Trager et al. (2000a) sample, 41 field galaxies are plotted in Figures 8(b), (d) and (f). A similar procedure has been used to convert the Trager et al. (2000a) parameters to our reference frame, using the relations Age$_{corr}$ = 0.95 $\times$ Age$_{Trager}$ + 1.76; [Z/H]$_{corr}$ = 0.27 $\times$ [Z/H]$_{Trager}$ + 0.15 and [Mg/Fe]$_{corr}$ = 1.05 $\times$ [Mg/Fe]$_{Trager}$ + 0.05. The resulting sample, comprising 40 field galaxies, is also presented in Figures 8(b), (d) and (f) where a common least-squares fits to all the data, including ours, have been displayed.

Proctor et al. (2004) and Mendes de Oliveira et al. (2005) studied elliptical galaxies in compact groups which can be used to assess our results. The first study is especially difficult to convert to our reference frame, because it follows a unique approach, very dissimilar to ours, and has only three galaxies in common. In addition, we suspect that their [Mg/Fe] results differ systematically from ours because their field galaxies do not show the striking [Mg/Fe] vs. log$\sigma_0$ correlation seen in our data and those of other authors, as displayed in Figure 8(f)). The comparison with Mendes de Oliveira et al. (2005) is simpler because they use the same TMB03 stellar population analysis. The conversion is carried out in two steps, starting with a transformation of their data from an aperture of R$_{eff}$/2 to R$_{eff}$/8, using the average parameter gradients obtained from our data, $\sigma_8$ = 1.1 $\times$ $\sigma_2$ -- 14.8. Because the reference frame of the Mendes de Oliveira et al. (2005) data is the G93 sample reduced with the TMB03 models, we use our twelve galaxies from the G93 sample to derive the conversions. The final relations are $\sigma_{corr}$ = 0.96 $\times$ $\sigma_{Mendes}$ + 24.3; Age$_{corr}$ = 1.04 $\times$ Age$_{Mendes}$ + 0.82; [Z/H]$_{corr}$ = 0.49 $\times$ [Z/H]$_{Mendes}$ + 0.04 and [Mg/Fe]$_{corr}$ = 0.94 $\times$ [Mg/Fe]$_{Mendes}$ + 0.05, with the results displayed in Figures 8(a), (c) and (e).

The conversion of the Mendes de Oliveira et al. (2005) data to our TMB03 frame adds a new object to the critical interval log$\sigma_0$ = 2.0 - 2.2 where the {\it anomaly} is detected. Their galaxy HCG76d, at log$\sigma_0$ = 2.19, shows the typical anomalous behavior, with enhanced [Mg/Fe] = 0.47 and reduced [Z/H] = +0.07.

\begin{figure}
\epsfxsize=9cm
\epsfbox{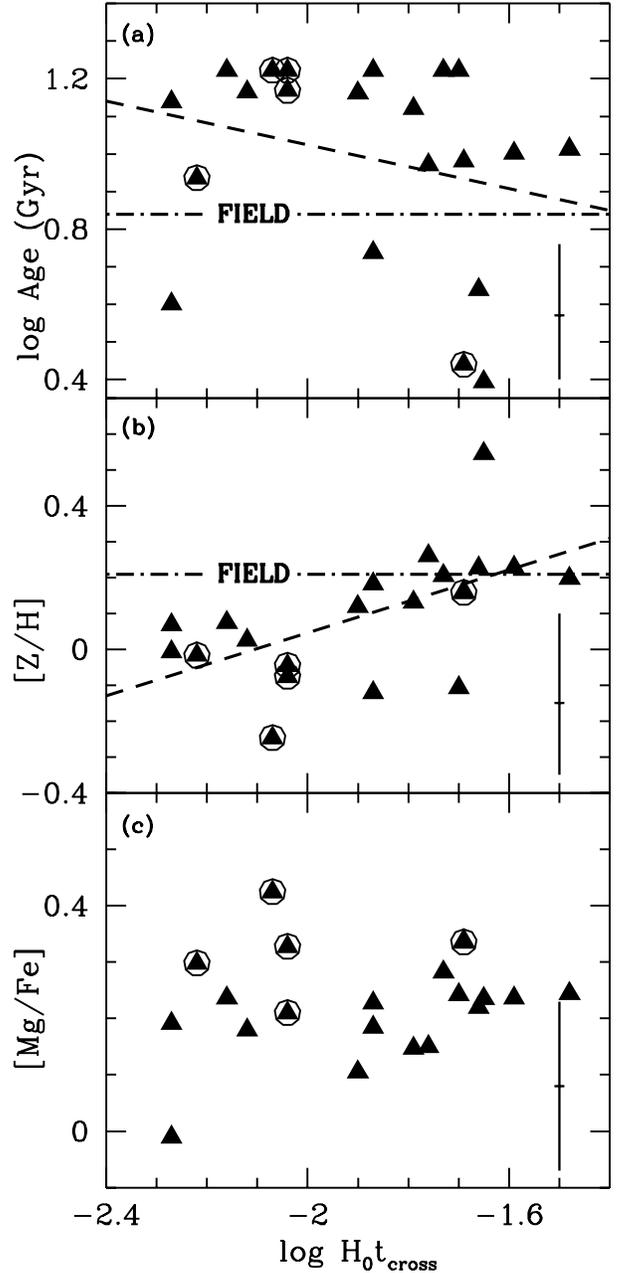}
\caption{Dependence of the parameters for galaxies in HCGs with the dynamical state of the group, represented by the {\it crossing time} log(H$_0$t$_c$). Least-squares fits (dashed lines) are displayed in panels (a) and (b), with the average value (dot-dashed lines) of the corresponding parameter in the field, except for panel (c), where no correlation has been found for [Mg/Fe]. {\it Anomalous} galaxies, defined in the text, have been identified with a circle.}
\end{figure}

\subsection{Stellar Population Parameters vs. Environment}
From our simplistic environmental division into HCGs and the field, we can conclude from Figure 8 that {\it our galaxies in the HCGs are older and more metal-poor than their counterparts in the field}, in agreement with previous works, e.g. de la Rosa et al. (2001b), Proctor et al. (2004) and Mendes de Oliveira et al. (2005). Our galaxies in groups are, on average, 1.6 Gyrs older than in the field. When field galaxies are discriminated according to their {\it local density}, those located in the less dense environments are the youngest. For instance, of the five oldest galaxies in Figure 8(b), four are located in {\it high density} environments, according to the classification of Thomas et al. (2005). Galaxies NGC4552, NGC4649 and NGC4697 are located in the Virgo cluster, while NGC7619 belongs to the Pegasus-I group. Would they have been excluded from the sample, the difference between groups and the field (low density) would increase to 4.3 Gyrs. The metallicity [Z/H] values for the field galaxies are 0.11 dex larger, on average, than those of the CGs, while the average [Mg/Fe] ratios are identical in both environments. Much more interesting environmental dependencies of the [Z/H] and [Mg/Fe] have been mentioned in the previous subsection, concerning their trends with log $\sigma_0$.

\input{tab3.tex}

From a dynamical point of view, HCGs form a heterogeneous sample, containing groups at different evolutionary stages (Ribeiro et al. 1998). A convenient parameter to characterize the degree of dynamical evolution of the group is the {\it crossing time} (H$_0$t$_c$), whose values we have taken from Hickson et al. (1992). For instance, these authors found a deficit of late-type galaxies in groups with shorter t$_c$, where galaxy-galaxy interactions are supposed to be more frequent.

In our study, Figures 9(a) and (b), we can see that both age and metallicity depend on the dynamical evolution of the HCGs. Least-squares fits are drawn to show that stellar population parameters for the less dynamically evolved HCGs, i.e. high log(H$_0$t$_c$), closely resemble those of the field, while the parameters of the HCGs with a small log(H$_0$t$_c$) show the largest disagreement with the field. This is a reasonable result, as it is generally accepted that galaxy interactions are the main driver of the environmental differences. The [Mg/Fe] does not show a clear dependence on the {\it crossing time}, (Figure 9(c)), but it is worth noticing that the {\it anomalous} galaxies, marked with circles, are preferentially located in the more evolved HCGs (log(H$_0$t$_c$) $\leq$ -2.0, suggesting that the source of the anomaly is related to the frequency of galaxy interactions. 

\subsection{Which are the "Anomalous" Galaxies?}

The {\it anomalous} galaxies in our sample are HCG14b, HCG46c, HCG59b, HCG19a, HCG46a and HCG57f, although the first three are the most extreme cases. From external samples, we can add to this list HCG76d (Mendes de Oliveira et al. 2005). Table 3 shows a comparison of the properties of the {\it anomalous} with the remainder of the HCG galaxies in our sample, excluding the dwarf HCG37e. As already mentioned, the {\it anomalous} galaxies show, on average, lower $\sigma_0$ (col. 2) and [Z/H] (col.3) as well as a larger [Mg/Fe] (col. 4) than the rest of the HCG sample. They are also located in more dynamically evolved groups, having shorter {\it crossing times} (col. 5). We can estimate the stellar masses of the galaxies based on the relation given by Thomas et al. (2005), $log M_{\star} \approx 0.63 + 4.52\ log \sigma$. The average mass of the {\it anomalous} galaxies thus calculated is 2.6$\times$10$^{10}$M$_{\sun}$, while for the rest of the sample it is 1.3$\times$10$^{11}$M$_{\sun}$.

\subsubsection {Activity}
Two thirds of the {\it anomalous} galaxies show nebular emission, against only one third of the rest of the HCG galaxies. We have studied their ionization mechanisms with the standard procedure, based on the characteristic spectral line ratios of the different activity classes (e.g. Coziol et al 1998). The emission lines were measured in the template-subtracted spectra, by following the same procedure described in section 3. According to the standard diagnostic diagram (Veilleux \& Osterbrock 1987), which uses the [NII]/H$\alpha$ and [OIII]/H$\beta$ line ratios, only one of the emission-line galaxies shows weak star formation activity, while the rest are classified as AGNs. All {\it active} galaxies increase their [OIII]$\lambda$5007 emission inwards, from $R_{eff}/2$ to $R_{eff}/8$, reinforcing the idea of a nuclear source of activity. 

\subsubsection {Photometric Profiles}
Mendes de Oliveira \& Hickson (1994) find that three of our {\it anomalous} galaxies, namely HCG 14b, 57f and 59b, stand out for having extended profiles and being more diffuse than other galaxies of the same luminosity. They note that these galaxies are similar to the cD galaxies in clusters, albeit much less luminous.Isophotal photometry for the galaxies in this sample (Ferreras et al. 2006) shows that all the {\it anomalous} galaxies with measured a4 parameters have disky isophotes (positive a4), unlike the rest of the sample, where 54\% of the galaxies have boxy profiles (negative a4). This can be a selection effect, as disky galaxies tend to be less luminous than their boxy counterparts,  although with a substantial amount of crossover (e.g. Bender et al. 1989).

\section{DISCUSSION}

Unlike the largest ellipticals, which are already inactive by z $\approx$ 1, medium-sized and dwarf galaxies in the field are still actively forming stars in the present universe. This phenomenon, known as "downsizing", has been repeatedly observed (Cowie et al. 1996; Kauffmann et al. 2003; Thomas et al, 2005) and recently interpreted in terms of a feedback-regulated picture of galaxy formation (Scannapieco, Silk \& Bouwens, 2005). Studies based on redshift surveys comprising vast samples of galaxies show a clear bimodality in star formation properties, with two distinct galaxy populations (Baldry et al. 2004). The {\it red sequence} comprises passive galaxies without SF, while the {\it blue sequence} includes galaxies with active SF. The global decrease observed in the SF rate density since redshift 1-2 (e.g. Hopkins \& Beacom 2006) suggests that galaxies are systematically transferred from the blue to the red sequence. Several observations can be used to outline the properties of the mechanism responsible for the transition or SF-truncation. The mere existence of the bimodality implies that the mechanism must be {\it fast} in order to preserve the separation of the sequences. A slow SF truncation mechanism is not viable, as it would smear out that separation. According to a study by Mateus et al. (2006), the galaxies filling the gap between both color sequences are mainly AGN hosts, indicating that the SF truncation mechanism must also involve {\it nuclear activity}. Concerning the preferred environment in which the mechanism might act, it has been repeatedly found that dense environments are hostile to SF (e.g. Baldry et al. 2004). However, extreme densities are not needed, because Lewis et al. (2002) and G\'omez et al. (2003) found that the SF truncation mechanism operates at the low local densities found well beyond the virialized cluster cores. Mateus \& Sodr\'e (2004) went further, showing that the SF in galaxies is affected by a wide range of environments, including galaxy groups and the field. A third characteristic of the SF truncation mechanism is that it must be able to {\it act everywhere}. Finally, Kauffmann et al. (2003) found that the transition between both sequences happens at a critical stellar mass 3$\times$10$^{10}$M$_{\sun}$, suggesting that this is the typical mass of the galaxies undergoing SF-truncation.

Although there is no shortage of mechanisms capable of truncating star formation in galaxies, none of them fulfills the three expected characteristics. For instance, {\it ram pressure stripping} (Dressler \& Gunn 1983) needs high velocities and densities at the inter-galactic medium, which are only found in dense clusters. The perturbation of the gas-rich halo of the galaxy, called {\it strangulation} (Larson, Tinsley \& Caldwell 1980) produces a too slow decrease of the SF, in $\geq$ 2 Gyrs (Balogh et al. 2004). Nevertheless, recent hydrodynamical simulations (Di Matteo, Springel \& Hernquist 2005) have shown that {\it galaxy mergers}, commonly considered a mechanism for SF enhancement, possesses powerful SF truncation properties. According to the simulations, the galaxy merger induces a short-lived star formation burst and the triggering of an AGN, which rapidly depletes the gas and quenches the star formation. Consequently, the merger of galaxies fulfills the three characteristics required for the SF truncation mechanism, i.e. it is fast, involves AGNs and can potentially act everywhere.
 
Compact groups of galaxies, with their high densities and low velocities make the ideal environment for mergers to occur. This line of reasoning lead us to concentrate our search for SF-truncation tracers in this minority class. Considering the plethora of faint galaxies detected in HCGs (de Carvalho et al. 1997), it is worth considering the possibility of minor mergers involving satellite or dwarf companions (Hernquist \& Mihos 1995; Bell et al. 2005). Galaxy-galaxy interactions, and especially minor mergers, are expected to occur in loose groups or even the field. In summary, we propose a scenario in which galaxies that enter the group environment suffer a minor merger. The consequences of the merger depend on the mass of the entering galaxy. For instance, due to the {\it downsizing}, massive galaxies have completed their SF long before entering the group and therefore only suffer mild consequences. A dwarf galaxy, with an extended SF history, will just contaminate the massive galaxy with [Z/H]-rich and [Mg/Fe]-poor stellar populations. However, a massive galaxy is passive and there is no SF to truncate. In contrast, a minor merger with a low-mass galaxy would likely induce more drastic consequences. To begin with, an unperturbed low-mass galaxy is expected to be still actively forming stars in the present universe. The merger event would deplete its gas and truncate any further SF, leaving the values of the stellar population parameters frozen. Unperturbed in the field, its counterparts with similar mass steadily increase their metallicity [Z/H] and decrease the [Mg/Fe], by incorporating Fe in the newly formed stars.  

Our observations fit this scenario well. First of all, we observe that galaxies in CGs are, on average, older and more metal-poor than their counterparts in the field, reminding us of the hostility to SF which characterizes dense environments. The anomalous enhanced [Mg/Fe] abundance ratio and deficient metallicity detected in the low-mass galaxies of our HCG sample, can be interpreted in the frame of the aforementioned {\it merger + SF-truncation} mechanism. We emphasize that both the [Mg/Fe] ratio and the [Z/H] are not constant but evolve along the SF history of the galaxy. Soon after the  SF history starts, the SN-II enrich the ISM with $\alpha$ elements which are readily incorporated into the newly forming stars. At that moment, the [Mg/Fe] ratio attains its highest value, which later decreases as soon as the Fe is released by the SNIa and incorporated into the stars. Similarly, along the SF history, the global metallicity [Z/H] of the stellar population increases. The SF-truncation puts an end to this evolution and leaves the galaxy with their parameter values at the time of the merger event.

The {\it anomalous} galaxies with larger [Mg/Fe] and lower [Z/H], are simply galaxies which truncated their SF. The fact that they belong to the densest groups (shorter T$_{cross}$), show larger incidence of AGN and a predominance of {\it disky} profiles reinforces the idea that their SF was truncated by a minor merger. Simulations show that disky ellipticals are the result of mergers between progenitors of disparate masses (Khochfar \& Burkert 2005; Jesseit, Naab \& Burkert 2005). The fact that our {\it anomalous} galaxies have an average stellar mass of 2.6$\times$10$^{10}$M$_{\sun}$ locates them in the transition region between the red and blue sequences defined by Kauffmann et al. (2003), where the SF-truncation is likely taking place. 

\section{SUMMARY AND CONCLUSIONS}

In this paper we have studied the stellar populations of elliptical galaxies in HCGs in order to compare them with their counterparts in the field. Assuming that galaxies in groups have been incorporated from the field, differences between members of those environments would unmask the long sought mechanism which makes the dense environments hostile to star formation. 

Our stellar population study has tried to minimize the sources of systematic effects by carrying out several parallel stellar population analyses with different procedures, models and emission corrections. Although the three alternative models provide different absolute values for the stellar population parameters, i.e. age, metallicity and abundance ratio, their results, concerning the differences between both field and HCG galaxies are qualitatively similar. 

From our results we can extract the following conclusions: 

(i) Galaxies in the HCGs are, on average, 1.6 Gyrs older and more metal-poor, in 0.11 dex of [Z/H], than their counterparts in the field. This result is in agreement with the previous findings reported in the literature and supports the view of a hostility of the dense environments to star formation.  

(ii) The [Mg/Fe] of galaxies in the field show a trend with their velocity dispersion ($\sigma_0$). More massive galaxies (larger $\sigma_0$) show larger abundance ratios. This result, also in agreement with previous studies, has been interpreted as a sign of {\it downsizing}, where the less massive galaxies have more extended SF histories.

(iii) The low-mass galaxies in HCGs (with average M$_{\star} \approx$ 2.6$\times$10$^{10}$M$_{\sun}$) show a distinct behavior with respect to similar galaxies in the field. Those in HCGs show higher [Mg/Fe] abundance ratios and lower [Z/H] metallicities. These {\it anomalous} galaxies also show other peculiarities when compared to the rest of the HCG sample. They are preferentially found in the most dynamically evolved HCGs (lower crossing times) and show an excess of AGNs, disky isophotes and extended photometric profiles, reminiscent of those in cD galaxies.    

We propose that low-mass galaxies from the field, actively forming stars, are incorporated into the dense CG environment, where they suffer mergers with other galaxies. As a result of the merger, the AGN feedback depletes the gas and quenches star formation. This SF truncation blocks the natural decrease of the [Mg/Fe] which accompanies any extended star formation. It equally hinders the metal enrichment and leaves a galaxy with larger [Mg/Fe] and lower [Z/H], as observed here. Massive galaxies, on the contrary, are excluded from this behavior because they are passive in the present universe and there is no SF to truncate. In summary, we conclude that we have detected traces of the merger action in low-mass galaxies of HCGs. 

If the {\it merger + SF-truncation} mechanism is accepted to be working in the Compact Groups, it can contribute to understanding a long standing apparent contradiction in HCGs, where their galaxies, despite showing abundant signs of tidal interactions, do not show the expected star formation enhancement. 

The {\it merger + SF-truncation} would possibly be the mechanism responsible for the global decay of the SF rate since redshift 1-2, as the groups are the places where the galaxies are pre-processed and transformed into passive objects, before being incorporated into clusters.

\acknowledgements
{We acknowledge Francesco Labarbera, Roy Gal and Paula Coelho for fruitful discussions along this project.}

\end{document}

%% file: tab1.tex
\begin{deluxetable}{lccccc}
\tabletypesize{\scriptsize}
\tablecaption{Basic Sample Parameters \label{T1}}
\tablecolumns{6}
\tablewidth{0pt}
\tablehead{
 Name & S/N  & $R_{eff}$  & $\theta$  & $\epsilon$ & Ap$_{R/8}$   \\
 & per \AA\ & arcsec & rad & & arcsec  
 }  
\startdata
\multicolumn{6}{c}{\bf COMPACT GROUP GALAXIES} \\[4 pt]
\hline \\  
HCG 10b &  53 & 11.4 & 0.35 & 0.16 &  3.0  \\ 
HCG 10b+& 121 & 11.4 & 2.90 & 0.16 &  3.0  \\	
HCG 14b &  25 & 47.6 & 0.52 & 0.41 & 12.8  \\	
HCG 15b &  39 &  8.0 & 1.85 & 0.10 & (1.9) \\	
HCG 15c &  41 & 10.1 & 1.40 & 0.12 &  2.4  \\	
HCG 19a &  55 &  9.0 & 2.44 & 0.32 &  2.2  \\  
HCG 19a+&  76 &  9.0 & 0.10 & 0.32 &  2.7  \\  
HCG 28b &  21 &  7.1 & 2.36 & 0.21 & (1.8) \\	
HCG 32a &  27 & 15.0 & 0.16 & 0.12 &  4.0  \\	
HCG 37a & 149 & 40.4 & 0.21 & 0.42 & 12.7  \\  
HCG 37e &  13 &  3.7 & 0.75 & 0.02 & (0.9) \\	
HCG 40a & 168 & 14.8 & 0.09 & 0.28 &  4.3  \\	
HCG 44b &  41 & 23.8 & 0.21 & 0.12 &  6.3  \\  
HCG 46a &  21 &(10.9)& 2.46 &(0.22)&  2.8  \\  
HCG 46c &  25 &  3.7 & 0.30 & 0.13 & (1.0) \\  
HCG 51a &  55 &(21.2)& 0.87 &(0.09)&  5.2  \\	
HCG 57c &  26 & 15.9 & 0.31 & 0.37 &  4.7  \\	
HCG 57c+&  59 & 15.9 & 2.95 & 0.37 &  4.9  \\	
HCG 57f &  60 & 10.2 & 2.78 & 0.39 &  3.0  \\	
HCG 59b &  23 &  6.6 & 1.57 & 0.10 & (1.6) \\  
HCG 62a &  46 &(42.0)& 3.00 &(0.29)& 12.3  \\  
HCG 68b &  31 & 32.0 & 0.10 & 0.00 &  8.0  \\
HCG 93a &  54 & 21.2 & 1.57 & 0.07 &  5.1  \\  
HCG 96b &  43 &  6.6 & 2.53 & 0.22 & (1.7) \\	
HCG 97a &  25 & 40.2 & 0.19 & 0.38 & 12.4  \\
\cutinhead{\bf FIELD GALAXIES}
NGC 221 & 255 & 39.0 & 0.17 & 0.00 &  9.8  \\  
NGC 584 &  67 & 30.0 & 1.05 & 0.30 &  6.7  \\  
NGC 584+&  70 & 30.0 & 1.83 & 0.30 &  6.4  \\  
NGC 636 &  94 & 19.0 & 0.70 & 0.13 &  4.8  \\  
NGC 821 &  62 & 36.0 & 2.71 & 0.32 &  9.9  \\  
NGC 1700 & 85 & 24.0 & 2.01 & 0.27 &  5.4  \\  
NGC 1700+& 69 & 24.0 & 2.79 & 0.27 &  6.7  \\  
NGC 1700++&47 & 24.0 & 1.41 & 0.27 &  5.1  \\  
NGC 2300&  70 & 34.0 & 1.78 & 0.16 &  7.8  \\  
NGC 3377& 123 & 34.0 & 2.53 & 0.50 &  8.5  \\  
NGC 3379&  60 & 35.0 & 1.90 & 0.09 &  8.4  \\  
NGC 4552&  71 & 30.0 & 2.36 & 0.07 &  7.5  \\
NGC 4649&  43 & 74.0 & 1.31 & 0.17 & 17.0  \\
NGC 4697& 105 & 75.0 & 1.92 & 0.41 & 15.0  \\  
NGC 7619&  53 & 32.0 & 2.62 & 0.24 &  8.4  \\
\enddata  
\end{deluxetable}

%% file: tab2.tex
\begin{deluxetable}{lrrrrrrrrrrrrrrr}
\small
\tablecaption{Extracted Parameters in R$_{eff}$/8 \label{T2}}
\tablecolumns{16}
\tabletypesize{\scriptsize}
\setlength{\tabcolsep}{0.05in}
\tablewidth{0pt}
\tablehead{
\colhead{} &  \colhead{} & \multicolumn{3}{c}{Vaz99} & \colhead{} & \multicolumn{3}{c}{TMB03} & \colhead{} & \multicolumn{2}{c}{MCGB05} & \colhead{} & \multicolumn{3}{c}{AVERAGE}\\
\cline{3-5} \cline{7-9} \cline{11-12} \cline{14-16}\\
\colhead{Name} & \colhead{$\sigma$(err)} & \colhead{Age} & \colhead{[Z/H]} & \colhead{[Mg/Fe]} &  \colhead{} & \colhead{Age} & \colhead{[Z/H]} & \colhead{[Mg/Fe]} & \colhead{} & \colhead{[Z/H]} & \colhead{[Mg/Fe]} & \colhead{} & \colhead{Age} & \colhead{[Z/H]} & \colhead{[Mg/Fe]}
}
\startdata
\multicolumn{16}{c}{\bf COMPACT GROUP GALAXIES}\\[4 pt]
\hline \\
HCG10b+  & 222.0( 2.7)& 12.2 & 0.01 & 0.53 & & 10.2 &  0.24 &  0.23 & & 0.18  &  0.33 & &10.3$^{+2.8}_{-2.7}$ &  0.20$^{+0.11}_{-0.10}$ &  0.24$^{+0.06}_{-0.06}$  \\[2pt]
HCG14b(*)& 141.5( 9.5)&  8.6 &-0.17 & 0.48 & & 10.0 & -0.07 &  0.38 & &-0.18  &  0.39 & & 8.6$^{+4.6}_{-4.8}$ & -0.02$^{+0.45}_{-0.42}$ &  0.30$^{+0.23}_{-0.24}$  \\[2pt]
HCG15b   & 171.8( 4.6)& 19.0 &-0.48 & 0.09 & & 17.0 & -0.11 &  0.22 & &-0.25  &  0.28 & &16.7$^{+6.5}_{-6.5}$ & -0.12$^{+0.17}_{-0.11}$ &  0.18$^{+0.14}_{-0.14}$  \\[2pt]
HCG15c   & 186.5( 8.8)&  7.1 &-0.01 & 0.16 & &  5.0 &  0.13 &  0.21 & & 0.32  &  0.52 & & 5.5$^{+6.3}_{-2.9}$ &  0.18$^{+0.32}_{-0.27}$ &  0.23$^{+0.16}_{-0.16}$  \\[2pt]
HCG19a+  & 161.4( 7.8)&  6.8 &-0.06 & 0.38 & &  5.0 &  0.08 &  0.22 & & 0.14  &  0.35 & & 5.3$^{+2.5}_{-1.8}$ &  0.12$^{+0.16}_{-0.14}$ &  0.23$^{+0.08}_{-0.08}$  \\[2pt]
HCG28b(*)& 173.1( 9.4)& 19.0 &-0.45 & 0.13 & & 17.0 & -0.12 &  0.30 & &-0.20  &  0.45 & &16.7$^{+8.0}_{-7.7}$ & -0.11$^{+0.37}_{-0.35}$ &  0.24$^{+0.27}_{-0.26}$  \\[2pt]
HCG32a   & 223.4( 5.1)&  5.7 &-0.02 & 0.21 & &  4.0 &  0.19 &  0.17 & & 0.46  &  0.52 & & 4.4$^{+5.9}_{-4.5}$ &  0.23$^{+0.32}_{-0.28}$ &  0.22$^{+0.20}_{-0.20}$  \\[2pt]
HCG37a(*)& 238.4( 7.3)& 16.4 &-0.12 & 0.36 & & 13.4 &  0.07 &  0.18 & &-0.04  &  0.22 & &13.7$^{+2.7}_{-2.6}$ &  0.07$^{+0.06}_{-0.06}$ &  0.19$^{+0.03}_{-0.03}$  \\[2pt]
HCG37e   &  86.0(28.3)&  3.6 &-0.21 &-0.03 & &  5.1 & -0.11 & -0.14 & & 0.00  & -0.10 & & 4.0$^{+3.4}_{-2.5}$ & -0.01$^{+0.50}_{-0.49}$ & -0.01$^{+0.33}_{-0.31}$  \\[2pt]
HCG40a   & 230.5( 7.6)& 16.8 &-0.18 & 0.23 & & 14.8 &  0.00 &  0.17 & &-0.07  &  0.26 & &14.6$^{+2.1}_{-1.9}$ &  0.02$^{+0.06}_{-0.05}$ &  0.18$^{+0.03}_{-0.03}$  \\[2pt]
HCG44b   & 169.2(24.3)&  3.1 & 0.58 & 0.42 & &  2.5 &  0.62 &  0.20 & & 0.60  &  0.41 & & 2.5$^{+3.5}_{-1.0}$ &  0.55$^{+0.27}_{-0.33}$ &  0.23$^{+0.15}_{-0.15}$  \\[2pt]
HCG46a(*)& 153.6(12.8)& 19.0 &-0.38 & 0.45 & & 17.0 & -0.02 &  0.26 & &-0.31  &  0.11 & &16.7$^{+8.1}_{-7.7}$ & -0.08$^{+0.39}_{-0.23}$ &  0.21$^{+0.25}_{-0.24}$  \\[2pt]
HCG46c(*)& 150.0(16.9)& 16.1 &-0.29 & 0.56 & & 15.7 & -0.11 &  0.35 & &-0.10  &  0.58 & &14.8$^{+7.0}_{-7.5}$ & -0.04$^{+0.33}_{-0.25}$ &  0.33$^{+0.20}_{-0.20}$  \\[2pt]
HCG51a   & 238.8( 9.6)& 19.0 & 0.07 & 0.59 & & 17.0 &  0.29 &  0.30 & & 0.05  &  0.38 & &16.7$^{+4.6}_{-4.5}$ &  0.21$^{+0.15}_{-0.15}$ &  0.28$^{+0.10}_{-0.09}$  \\[2pt]
HCG57c+  & 213.7( 5.7)& 10.3 & 0.03 & 0.71 & & 10.4 &  0.16 &  0.36 & & 0.10  &  0.52 & & 9.6$^{+3.9}_{-3.1}$ &  0.16$^{+0.17}_{-0.15}$ &  0.34$^{+0.09}_{-0.09}$  \\[2pt]
HCG57f   & 160.8(26.6)&  3.4 &-0.03 & 0.72 & &  2.8 &  0.10 &  0.31 & & 0.27  &  0.60 & & 2.8$^{+1.3}_{-0.8}$ &  0.16$^{+0.17}_{-0.13}$ &  0.34$^{+0.09}_{-0.09}$  \\[2pt]
HCG59b(*)& 112.1(25.6)& 19.0 &-0.54 & 0.66 & & 17.0 & -0.25 &  0.67 & &-0.60  &  0.45 & &16.7$^{+7.7}_{-7.5}$ & -0.25$^{+0.35}_{-0.29}$ &  0.42$^{+0.26}_{-0.26}$ \\[2pt]
HCG62a(*)& 230.0( 3.1)& 19.0 &-0.13 & 0.50 & & 17.0 &  0.13 &  0.25 & &-0.10  &  0.26 & &16.7$^{+5.2}_{-4.9}$ &  0.08$^{+0.19}_{-0.19}$ &  0.24$^{+0.13}_{-0.14}$  \\[2pt]
HCG68b(*)& 203.0(23.3)& 15.5 &-0.05 & 0.41 & & 13.1 &  0.14 &  0.05 & & 0.08  &  0.18 & &13.2$^{+6.5}_{-6.2}$ &  0.13$^{+0.31}_{-0.22}$ &  0.15$^{+0.17}_{-0.17}$  \\[2pt]
HCG93a(*)& 220.7( 5.3)& 11.5 & 0.13 & 0.57 & & 10.3 &  0.28 &  0.23 & & 0.11  &  0.25 & &10.0$^{+4.4}_{-4.1}$ &  0.23$^{+0.17}_{-0.19}$ &  0.24$^{+0.11}_{-0.11}$  \\[2pt]
HCG96b   & 200.2(16.5)& 12.8 & 0.04 & 0.20 & &  7.8 &  0.35 &  0.11 & & 0.28  &  0.22 & & 9.4$^{+5.0}_{-3.8}$ &  0.26$^{+0.24}_{-0.23}$ &  0.15$^{+0.13}_{-0.13}$  \\[2pt]
HCG97a   & 189.5( 4.4)& 19.1 &-0.09 & 0.21 & & 12.6 &  0.17 &  0.04 & & 0.01  &  0.08 & &14.5$^{+6.8}_{-6.8}$ &  0.12$^{+0.34}_{-0.27}$ &  0.11$^{+0.15}_{-0.15}$  \\
\cutinhead{\bf FIELD GALAXIES}
NGC221   &  80.4(14.8)&  3.5 & 0.00 &-0.37 & &  2.5 &  0.15 & -0.03 & & 0.00  & -0.20 & & 2.6$^{+0.2}_{-0.2}$ &  0.13$^{+0.04}_{-0.04}$ & -0.03$^{+0.02}_{-0.02}$  \\[2pt]
NGC584+(*)& 179.6(10.3)&  6.5 & 0.19 & 0.28 & &  4.4 &  0.29 &  0.24 & & 0.55  &  0.48 & & 4.9$^{+2.5}_{-2.3}$ &  0.33$^{+0.14}_{-0.14}$ &  0.24$^{+0.06}_{-0.07}$  \\[2pt]
NGC636   & 161.8( 7.6)&  8.9 & 0.10 & 0.41 & &  4.7 &  0.20 &  0.16 & & 0.30  &  0.41 & & 6.1$^{+1.9}_{-1.0}$ &  0.23$^{+0.11}_{-0.11}$ &  0.22$^{+0.05}_{-0.06}$  \\[2pt]
NGC821   & 184.3( 1.9)&  9.5 & 0.23 & 0.65 & &  7.6 &  0.32 &  0.26 & & 0.33  &  0.50 & & 7.8$^{+3.1}_{-2.2}$ &  0.31$^{+0.14}_{-0.15}$ &  0.29$^{+0.07}_{-0.07}$  \\[2pt]
NGC1700  & 220.3(10.3)&  7.9 & 0.18 & 0.43 & &  5.3 &  0.27 &  0.15 & & 0.40  &  0.40 & & 6.0$^{+2.1}_{-1.2}$ &  0.29$^{+0.11}_{-0.12}$ &  0.22$^{+0.05}_{-0.05}$  \\[2pt]
NGC2300  & 245.3(14.9)&  7.1 & 0.20 & 0.72 & &  5.0 &  0.40 &  0.38 & & 0.70  &  0.80 & & 5.5$^{+1.6}_{-1.4}$ &  0.40$^{+0.11}_{-0.10}$ &  0.39$^{+0.05}_{-0.05}$  \\[2pt]
NGC3377  & 142.5( 8.9)&  6.9 & 0.02 & 0.35 & &  6.2 &  0.11 &  0.21 & & 0.18  &  0.38 & & 6.0$^{+1.1}_{-0.9}$ &  0.16$^{+0.05}_{-0.06}$ &  0.23$^{+0.03}_{-0.03}$  \\[2pt]
NGC3379  & 182.5( 5.9)& 19.3 &-0.14 & 0.56 & & 15.5 &  0.04 &  0.27 & &-0.06  &  0.40 & &16.1$^{+2.6}_{-2.7}$ &  0.05$^{+0.07}_{-0.08}$ &  0.27$^{+0.06}_{-0.06}$  \\[2pt]
NGC4552  & 240.1( 8.8)& 19.8 &-0.11 & 0.59 & & 17.0 &  0.20 &  0.33 & & 0.11  &  0.53 & &17.0$^{+4.7}_{-4.8}$ &  0.14$^{+0.18}_{-0.16}$ &  0.32$^{+0.06}_{-0.06}$  \\[2pt]
NGC4649  & 283.5(13.3)& 19.0 &-0.11 & 0.70 & & 17.0 &  0.37 &  0.26 & & 0.18  &  0.39 & &16.7$^{+4.9}_{-4.7}$ &  0.21$^{+0.17}_{-0.19}$ &  0.28$^{+0.10}_{-0.10}$  \\[2pt]
NGC4697  & 155.4(24.8)& 13.3 &-0.01 & 0.26 & & 13.4 &  0.09 &  0.14 & & 0.00  &  0.23 & &12.4$^{+2.0}_{-1.9}$ &  0.11$^{+0.10}_{-0.06}$ &  0.17$^{+0.04}_{-0.04}$  \\[2pt]
NGC7619  & 298.6( 2.2)& 14.6 &-0.05 & 0.81 & & 14.6 &  0.00 &  0.25 & & 0.08  &  0.58 & &13.6$^{+3.8}_{-4.0}$ &  0.09$^{+0.18}_{-0.10}$ &  0.32$^{+0.09}_{-0.09}$  \\
\enddata
\end{deluxetable}

%% file: tab3.tex
\begin{deluxetable*}{lcccccc}
\footnotesize
\tablecaption{Peculiarities of the {\it anomalous} galaxies\label{T3}}
\tablewidth{0pt}
\tablehead{
Subsample & $\sigma_0$ & [Z/H] & [Mg/Fe] & log(H$_0$t$_c$) & Emission & a4$\geq$0 \\
}
\startdata
Anomalous & 146.6 $\pm$ 18.4 &  -0.02 $\pm$ 0.15 & 0.30 $\pm$ 0.08 & -2.01 $\pm$ 0.19 & 67 \% & 100 \% \\[2pt]
Rest      & 207.4 $\pm$ 24.5 &  0.15 $\pm$ 0.16 & 0.21 $\pm$ 0.06 & -1.82 $\pm$ 0.22 & 33 \% & 46 \% \\[2pt]
\enddata  
\end{deluxetable*}